\documentclass[a4paper,11pt]{article}
\pdfoutput=1 

\usepackage{jheppub} 

\usepackage[T1]{fontenc} 

\usepackage{tikz}
\usetikzlibrary{decorations.pathmorphing}
\usepackage{capt-of}
\usepackage{caption}
\usepackage{subcaption}

\usepackage{amssymb}
\usepackage{epsfig, float}

\newcommand{\be}{\begin{equation}}
\newcommand{\ee}{\end{equation}}
\newcommand{\bea}{\begin{eqnarray}}
\newcommand{\eea}{\end{eqnarray}}

\newcommand{\mt}[1]{\textrm{\tiny #1}}

\newcommand{\ket}[1]{|#1\rangle}

\def\uh {u_\mt{H}}

\newcommand*{\medcup}{\mathbin{\scalebox{1.15}[1.4]{\ensuremath{\cup}}}}

\title{\boldmath Delocalizing Entanglement \\
of Anisotropic Black Branes}

 \author{Viktor Jahnke}
 \affiliation{Departamento de F\'isica de Altas Energias, Instituto de Ciencias Nucleares, Universidad Nacional Aut\' onoma de M\' exico,\\ Apartado Postal 70-543, CDMX 04510, M\'exico}

\emailAdd{viktor.jahnke@correo.nucleares.unam.mx}

\abstract{We study the mutual information between pairs of regions on the two asymptotic boundaries of maximally extended anisotropic black branes. This quantity characterizes the local pattern of entanglement of the thermofield double states which are dual to these geometries. We analyze the disruption of the mutual information in anisotropic shock wave geometries and show that the entanglement velocity plays an important role in this phenomenon. Moreover, we compute several chaos-related properties of this system, such as the entanglement velocity, the butterfly velocity, and the scrambling time. We find that the butterfly velocity and the entanglement velocity violate the upper bounds proposed in \cite{tsunami1,tsunami2,Mezei-2016}, but remain bounded by their corresponding values in the infrared effective theory.}

\begin{document} 
\maketitle
\flushbottom

\section{Introduction} 
\label{sec-1}

In recent years, the chaotic properties of many-body quantum systems have started playing an important role in the search for a deeper understanding of the inner workings of the gauge/gravity duality \cite{duality1,duality2,duality3}. For example, the saturation of the so-called {\it chaos bound} might be a necessary condition for a large-$N$ system to have a gravitational dual \cite{bound-chaos,Kitaev2014}. On a more practical level, having in mind applications of the gauge/gravity duality to the study of strongly coupled systems, there seems to be an interesting connection between chaos and transport properties of such systems \cite{Blake1,Blake2}.

The hallmark of chaos is the sensitive dependence of evolution on initial conditions or, in a slogan, the {\it butterfly effect}. In the context of many-body quantum systems, this effect can be characterized by the commutator $[W(t,x),V(0)]$ between local hermitian operators $W$ and $V$. This commutator measures the influence of an early perturbation $V(0)$ on a later measurement of $W(t,x)$. The strength of the butterfly effect is usually characterized by
\be
C(t,x)= - \langle [W(x,t),V(0)]^2 \rangle_{\beta}\,,
\label{eq-C}
\ee
where $\langle \cdot \rangle_{\beta} = Z^{-1} \text{tr}\,[e^{\beta H} \cdot]$ denotes the thermal expectation value at the inverse temperature $\beta$. A physical system is said to be chaotic if $C(t,x) \approx 2 \,\langle W\,W \rangle \, \langle V\,V \rangle$ at large times for almost any choice of $W$ and $V$ \cite{AMPSS}. The time scale at which this occurs characterizes how fast the system can scramble information and is known as the {\it scrambling time} \cite{scrambling1,scrambling2}.

In the context of the gauge/gravity duality, the chaotic properties of the boundary theory at finite temperature can be calculated holographically by studying shock waves geometries in the bulk \cite{BHchaos1,BHchaos2,BHchaos3,BHchaos4}. For large-$N$ gauge theories this holographic approach gives
\be 
C(t,x) = \frac{K}{N^2} e^{\lambda_L (t-t_*-x/v_\mt{B})}+O(N^{-4})\,,\,\,\,\,\text{for}\,\,\,\, t \lesssim t_*\,,
\ee
where $t_*$ is the scrambling time, $\lambda_L$ is the Lyapunov exponent, and $v_\mt{B}$ is the butterfly velocity. The Lyapunov exponent characterizes the rate of growth of chaos and it is bounded by the temperature $\lambda_L \leq 2\pi/\beta$ \cite{bound-chaos}. As the Lyapunov exponent associated to black holes is always maximal, large-$N$ systems that saturate the chaos bound were conjectured to have an Einstein gravity dual \cite{bound-chaos,Kitaev2014}. The butterfly velocity characterizes the speed at which the perturbation $V$ grows. Together with the scrambling time, the butterfly velocity defines a {\it butterfly effect cone}, defined by $t-t_* = x/v_\mt{B}$. Inside of this cone, for $t-t_* \leq x/v_\mt{B}$, one expects to have $C(t,x)> \langle W\,W \rangle \, \langle V\,V \rangle$, whereas outside of the cone, for $t-t_* \geq x/v_\mt{B}$, one expects to have $C(t,x) \approx 0$.

In order to study chaos using holographic methods, one usually considers a thermofield double state of two identical copies of the boundary theory. Let us call them the left $L$ and right $R$ boundary theories. At $t=0$, the thermofield double state is given by
\be
\ket{TFD}  = \frac{1}{Z^{1/2}}\sum_{n} e^{-\frac{\beta}{2} E_{n}} \ket{n}_L \ket{n}_R\,.
\label{TFDstate}
\ee
For sufficiently high temperatures, the gravity dual of this system is a two-sided black hole \cite{eternalBH}. This is a wormhole geometry with two asymptotic AdS regions. The $L$ and $R$ theories are completely decoupled and live, respectively, on the left and right asymptotic boundaries of the geometry. The wormhole connecting the two sides of the geometry is not traversable. This is consistent with the fact that the two boundary theories are decoupled. The total amount of entropy between the two sides of the geometry is given by the Bekenstein-Hawking entropy of the black hole, which is equal to the cross sectional area of the wormhole.

At $t=0$, the thermofield double state (\ref{TFDstate}) displays local correlations between subsystems of $L$ and $R$, which signalize a highly atypical left-right entanglement pattern \cite{BHchaos1,BHchaos2,ER-EPR}. The butterfly effect can be characterized by the disruption of this local pattern of entanglement when a small perturbation is added to the system in the asymptotic past \cite{BHchaos1}. From the point of view of the boundary theory, this perturbation scrambles the left side Hilbert space, whereas, from the point of view of the bulk theory, a small perturbation applied early enough gets blue-shifted as it falls into the black-hole, generating a shock wave geometry. In both perspectives, the perturbed system no longer has two-sided local correlations at $t=0$\footnote{We emphasize that only two-sided local correlations are lost, not the local correlations within each side.}. 

The disruption of the local pattern of entanglement can be characterized by the mutual information $I(A,B)$ between two regions $A$ and $B$, located on the $t=0$ slice of the left and right boundaries, respectively. This quantity is given by
\be
I(A,B) = S(A)+S(B)- S(A \cup B)\,,
\ee
where $S(A)$ is the entanglement entropy of the region $A$, and so on. This quantity is always positive and it provides an upper bound for correlations between operators defined on the regions $A$ and $B$ \cite{boundIAB}. The above entanglement entropies can be calculated in holography using the prescriptions in \cite{RT,HRT}. The extremal surfaces which are homologous to $A$ ($B$) lie on the left (right) side of the geometry outside of the black hole horizon. The extremal surface homologous to $A \cup B$ can be given by the union of the two extremal surfaces homologous to $A$ and $B$, or by the surface that stretches through the wormhole connecting the two boundaries, depending on which one is smaller. If the two extremal surfaces homologous to $A$ and $B$ are smaller than the surface that connects the two boundaries, then the mutual information between $A$ and $B$ will be zero. If, on the other hand, the surface that stretches through the wormhole is the smaller one, then the mutual information will have some positive value.

For an appropriate choice of regions $A$ and $B$, the unperturbed geometry has $I(A,B)>0$.  When a small perturbation is added in the past, the blue-shift relative to the $t=0$ frame gives rise to a shock wave geometry in which the wormhole becomes longer. In this case, the extremal surface homologous to $A \cup B$ will also becomes longer, and the mutual information $I(A,B)$ will decrease. As we move the perturbation further into the past, the mutual information will eventually drop to zero,\footnote{Actually, as the Ryu-Takayanagi prescription only gives the leading order contribution to the entanglement entropy at large-$N$, only the leading contribution $I(A,B)$ will drop to zero.} characterizing the disruption of the local pattern of entanglement displayed by the unperturbed system at $t=0$. This is the holographic realization of the butterfly effect \cite{BHchaos1}.

In this paper, we study the disruption of the local pattern of entanglement in two-sided anisotropic black brane solutions.\footnote{This was first done for BTZ black holes in \cite{BHchaos1} and later generalized to higher dimensional cases in \cite{Leichenauer-2014} and to more general backgrounds in \cite{Sircar-2016}. Other works in this direction, include, for instance \cite{Huang-2016,Cai-2017,deBoer-2017,Murata-2017}.} In particular, we consider the anisotropic black brane solution of Mateos and Trancanelli  (MT) \cite{MT1,MT2}. By studying the shock wave geometries, we also find how the anisotropy affects other chaotic properties of these systems, like the Lyapunov exponent, the scrambling time, the butterfly velocity, and the entanglement velocity.

The paper is organized as follows. In section \ref{sec-2}, we review how to find consistent shock wave solutions for a very general five-dimensional anisotropic background. In that section we also review how to extract the Lyapunov exponent, the scrambling time, and the butterfly velocity from the profile of the shock wave solution. In section \ref{sec-3}, we compute the mutual information for strip-like regions both in the unperturbed geometry and in the shock wave geometry. We also explain the role of the entanglement velocity \cite{tsunami1,tsunami2,HM} in the disruption of the mutual information in shock wave geometries. We specialize our formulas for the anisotropic black brane MT solution and present several chaotic properties of this model. In particular, we show that both the butterfly velocity and the entanglement velocity violate the upper bounds proposed in \cite{tsunami1,tsunami2,Mezei-2016}, but remain bounded by their values in the infrared effective theory. Finally, we discuss our results in section \ref{sec-4}. We relegate to the Appendices \ref{appA} and \ref{appB} some technical details of the computation and a review the anisotropic MT model in appendix \ref{appC}.

\section{Gravity setup} 
\label{sec-2}

\subsection{Unperturbed geometry}
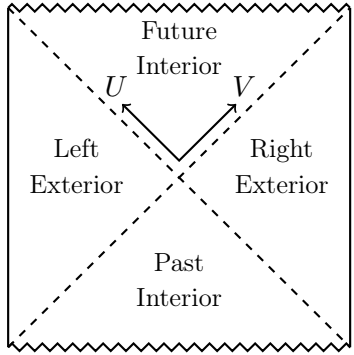
\begin{figure}[h!]
\centering
\captionsetup{justification=centering}

\begin{tikzpicture}[scale=1.5]
\draw [thick]  (0,0) -- (0,3);
\draw [thick]  (3,0) -- (3,3);
\draw [thick,dashed]  (0,0) -- (3,3);
\draw [thick,dashed]  (0,3) -- (3,0);
\draw [thick,decorate,decoration={zigzag,segment length=2mm, amplitude=0.5mm}]  (0,3) -- (3,3);
\draw [thick,decorate,decoration={zigzag,segment length=2mm,amplitude=.5mm}]  (0,0) -- (3,0);
\draw [thick,<->] (1,2.15) -- (1.5,1.65) -- (2,2.15);
\node [above] at (0.95,2.13) {$U$};
\node [above] at (2.07,2.13) {$V$};

\node[scale=0.9, align=center] at (1.5,2.65) {Future\\ Interior};
\node[scale=0.9,align=center] at (1.5,.6) {Past\\ Interior};
\node[scale=0.9,align=center] at (0.6,1.6) {Left\\ Exterior};
\node[scale=0.9,align=center] at (2.4,1.6) {Right\\ Exterior};
\end{tikzpicture}
\vspace{0.1cm}
\caption{ \small Penrose diagram for the two-sided black branes we consider.}
\label{fig-PenroseAdS}
\end{figure}

We consider a general 5-dimensional anisotropic metric of the form
\be
ds^2=G_{mn}dx^m dx^n=-G_{tt}(u)dt^2+G_{uu}(u)du^2+G_{xx}(u) \big(dx^2+dy^2 \big)+G_{zz}(u)dz^2\,.
\label{metric}
\ee
We take the boundary to be located at $u=0$, where the above metric is assumed to asymptote to $AdS_5$. We call $z$ the anisotropic direction and $x$ and $y$  the transverse directions. The horizon is located at $u=\uh$ where $G_{tt}$ has a first order zero and $G_{uu}$ has a first order pole. The other metric functions are assumed to be finite at the horizon. Near the horizon, the metric functions $G_{tt}$ and $G_{uu}$ can be written as
\be
G_{tt} = c_0 (u-\uh)\,,\qquad G_{uu} = \frac{c_1}{(u-\uh)}\,.
\label{GnearHorizon}
\ee
By requiring regularity of the Euclidean continuation of the above metric at the horizon, one obtains the inverse Hawking temperature as
\be
\beta = 4 \pi \sqrt{ \frac{c_1}{c_0} }\,.
\ee
In order to study shock waves geometries, it will be more convenient to work with Kruskal coordinates, since these coordinates cover smoothly the two sides of the geometry. We first introduce the Tortoise coordinate $u_*$ as
\be
du_{*} =-\sqrt{\frac{G_{uu}}{G_{tt}}}\, du \qquad  \text{or} \qquad u_{*}=-\int_0^{u} du' \sqrt{\frac{G_{uu}(u')}{G_{tt}(u')}} \,,
\ee
and then we define Kruskal coordinates $U,V$ in the left exterior region as
\bea
UV = -e^{\frac{4 \pi}{\beta} u_{*}}\label{eq-kruskalUV}\,,\qquad 
U/V =-e^{-\frac{4 \pi}{\beta} t} \label{eq-kruskalUsobV}\,.
\eea
In terms of these coordinates the metric can be written as
\be
ds^2=2 A(U,V) dU dV+ G_{ij}(U,V)dx^i dx^j\,,
\label{metricKruskal}
\ee
where
\be
A(U,V) = \frac{\beta^2 G_{tt}(U,V)}{8 \pi^2}\frac{1}{U V}\,,
\ee
and
\be
G_{ij}(U,V)dx^i dx^j =G_{xx}(U,V) \big(dx^2+dy^2 \big)+G_{zz}(U,V)dz^2\,.
\ee
In these coordinates the horizon is located at $U=0$ or $V=0$. The region $U>0$ and $V<0$ ($U<0$ and $V>0$) covers the left (right) exterior region. The boundary is located at $UV=-1$ and the black hole singularity at $UV=1$. The corresponding Penrose diagram is shown in figure \ref{fig-PenroseAdS}.

\subsection{Shock wave geometry}

In this section we explain how the unperturbed metric (\ref{metricKruskal}) is modified when a small pulse of energy is added from the boundary to the horizon in the left side of the geometry. Contrary to what one would naively assume, this perturbation has a non-trivial effect on the geometry \cite{BHchaos1}. This happens because, relative to the $t=0$ frame, the energy of the perturbation released a time $t_0$ in the past increases exponentially with $t_0$. For sufficiently large $t_0$, this blue-shift becomes so large that the perturbation follows an almost null trajectory close to the past horizon, giving rise to a shock geometry in this frame. Here we explicitly derive the shock wave solution for an anisotropic metric of the form (\ref{metricKruskal}). What follows is a review of the general analysis of \cite{Sfetsos-94} with appropriate modifications for an anisotropic background.\footnote{Anisotropic shock wave solutions were also studied in detail in \cite{Lifshitz-sw} for Lifshitz-like spacetimes.} We also review how the chaotic properties like the scrambling time, the butterfly velocity and the Lyapunov exponent can be extracted from the shock wave profile.

We assume the unperturbed metric (\ref{metricKruskal}) is a solution of Einstein's equations with energy-momentum tensor given by
\be
T^\mt{matter}_0= 2 T_{U V} dU dV + T_{U U} dU^2+T_{VV} dV^2+T_{ij}dx^i dx^j\,,
\ee
where $T_{mn} = T_{mn}(U,V,x^i)$. This is the most general energy momentum-tensor consistent with the Ricci tensor for unperturbed geometry.

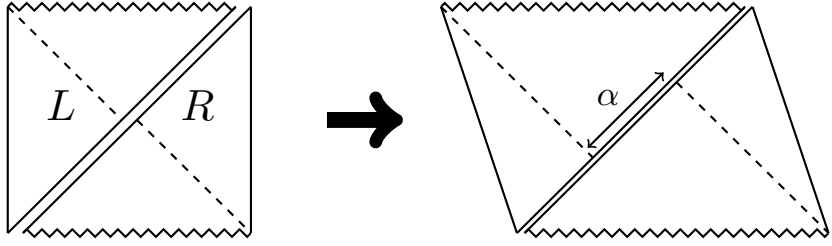
\begin{figure}[h!]
\centering
\captionsetup{justification=centering}

\begin{tikzpicture}[scale=1.]
\draw [thick]  (-0.2,0) -- (-0.2,3);
\draw [thick]  (3,0) -- (3,3);
\draw [thick]  (0,0) -- (3,3);
\draw [thick]  (-0.2,0) -- (2.8,3);
\draw [thick,dashed]  (-0.2,3) -- (1.3,1.5);
\draw [thick,dashed]  (1.5,1.5) -- (3,0);
\draw [thick,decorate,decoration={zigzag,segment length=2mm, amplitude=0.5mm}]  (-0.2,3) -- (2.8,3);
\draw [thick,decorate,decoration={zigzag,segment length=2mm,amplitude=.5mm}]  (0,0) -- (3,0);

\node[scale=1.5] at (.5,1.7) {$L$};
\node[scale=1.5] at (2.3,1.7) {$R$};

\draw [line width=6,->] (4,1.5) -- (5,1.5);

\draw [thick,decorate,decoration={zigzag,segment length=2mm, amplitude=0.5mm}]  (5.5,3) -- (9.5,3);

\draw [thick,decorate,decoration={zigzag,segment length=2mm, amplitude=0.5mm}]  (6.6,0) -- (10.6,0);

\draw [thick] (6.5,0) -- (9.5,3);
\draw [thick] (6.6,0) -- (9.6,3);

\draw [thick] (5.5,3) -- (6.5,0);

\draw [thick] (9.6,3) -- (10.6,0);

\draw [thick,dashed] (5.5,3) -- (7.5,1);
\draw [thick,dashed] (8.6,2) -- (10.6,0);

\draw [thick,<->] (8.43,2.13) -- (7.43,1.13);

\node [scale=1.2] at (7.7,1.8) {$\alpha$};

\end{tikzpicture}
\vspace{0.1cm}
\caption{ \small Penrose diagram for the shock wave geometry.}
\label{fig-PenroseShockWave}
\end{figure}

We want to know how the metric (\ref{metricKruskal}) changes when we add to the system a null pulse of energy located at $U=0$ and moving with the speed of light in the $V-$ direction.
The pulse worldline divides the spacetime into two regions, $L$ and $R$, as shown in the Penrose diagram in figure \ref{fig-PenroseShockWave}. The left region $L$ ($U>0$) is the causal future of the pulse, while the right region $R$ ($U<0$) is its causal past. The metric in the region $R$ should be the same as the unperturbed metric (\ref{metricKruskal}), whereas the metric in the region $L$ must be modified in order to account for the presence of the pulse of energy.

In isotropic geometries the back reaction of this pulse of energy in the left side of the geometry is simply obtained as a shift $V \rightarrow V + \alpha$ in the $V-$coordinate, where $\alpha=\alpha(t,x^i)$ is a function that can be determined from Einstein's equations \cite{Sfetsos-94,Dray-85}. To find shock wave solutions in anisotropic backgrounds we use the same anzats, which can be incorporated by replacing $V$ by $V+\theta(U) \alpha$ in the unperturbed metric (\ref{metricKruskal}). Note that the Heaviside's step function $\theta(U)$ ensures that the metric only changes in the causal future of the pulse (region $L$), remaining unchanged in the causal past (region $R$). The Penrose diagram of the corresponding shock wave geometry is shown in figure \ref{fig-PenroseShockWave}. 
With the ansatz above, the shock wave geometry is simply given by
\be
ds^2=2 A(U,V+\theta \alpha)\, dU \,(dV+\theta\, \partial_i\alpha\, dx^i)+
G_{ij}(U,V+\theta \alpha) dx^i dx^j\,,
\ee
while the energy-momentum of the matter fields is given by
\bea
\nonumber
T^\mt{matter}= 2\, T_{U V}(U,V+\theta \alpha) dU (dV+\theta\, \partial_i\alpha\, dx^i) + T_{U U}(U,V+\theta \alpha) dU^2\\ 
+T_{VV}(U,V+\theta \alpha) dV^2+T_{ij}(U,V+\theta \alpha) dx^i dx^j\,.
\eea
For convenience, we define new coordinates
\bea
\hat{U} = U\,,\qquad
\hat{V} =V+\theta(U)\,,\qquad
\hat{x}^i= x^i
\,,
\label{new-cordinates}
\eea
in terms of which the metric and the energy momentum tensor can be written as
\be
ds^2=2 \hat{A} \,d\hat{U} d\hat{V}+ \hat{G}_{ij}\, d\hat{x}^i d\hat{x}^j-2 \hat{A} \, \hat{\alpha} \, \delta(\hat{U}) \,d\hat{U}^2
\label{metric-ansatz}
\ee
and
\bea
\nonumber
T^\mt{matter}= 2\,\big[ \hat{T}_{\hat{U} \hat{V}}-\hat{T}_{\hat{V} \hat{V}}\, \hat{\alpha} \,\delta(\hat{U}) \big]d\hat{U} d\hat{V}
+\hat{T}_{\hat{V} \hat{V}} d\hat{V}^2+\hat{T}_{ij}d\hat{x}^i d\hat{x}^j+\\ 
\big[ \hat{T}_{\hat{U} \hat{U}}+\hat{T}_{\hat{V} \hat{V}}\, \hat{\alpha}^2\,\delta(\hat{U})^2-2\hat{T}_{\hat{U} \hat{V}} \hat{\alpha} \,\delta(\hat{U}) \big]d\hat{U}^2\,,
\label{Tmatter}
\eea
where the hats above $A$, $T_{mn}$ and $G_{ij}$ indicate that these quantities are calculated at $(\hat{U},\hat{V},\hat{x}^i)$.

The energy-momentum tensor of the pulse of energy that gives rise to the shock wave geometry is assumed to have the following form
\be
T^\mt{shock} = E\, e^{2 \pi t / \beta}\, \delta(\hat{U})\, a(\hat{x}^i) d\hat{U}^2\,,
\label{Tshock}
\ee
where $E$ is a constant related to the asymptotic energy of the pulse and $a(\hat{x}^i)$ specifies how localized is the perturbation. For a homogeneous perturbation, we take $a(\hat{x}^i)=1$, whereas for a localized perturbation we assume $a(\hat{x}^i)=\delta(\hat{x}^i)$. We want to find the function $\alpha(t,x^i)$ such that the ansatz (\ref{metric-ansatz}) satifies the Einstein's equations
\be
R_{mn}-\frac{1}{2}G_{mn} R = 8 \pi G_\mt{N} ( T^\mt{matter}_{mn}+T^\mt{shock}_{mn})
\,,
\label{EE}
\ee
with $T^\mt{matter}$ and $T^\mt{shock}$ given by (\ref{Tmatter}) and (\ref{Tshock}), respectively. In order to analyze the Einstein's equations it is convenient to rescale $\alpha$ and  $T^\mt{shock}$ as $\alpha \rightarrow \epsilon \, \alpha$ and $T^\mt{shock} \rightarrow \epsilon \, T^\mt{shock}$. By doing this we can recover the equations of motion for the unperturbed metric by setting $\epsilon = 0$ in  (\ref{EE}). In what follows we drop the hat over the symbols to simplify the notation, but one should remember that we are really using the new coordinates defined in (\ref{new-cordinates}). Assuming the Einstein's equations are satisfied for $\epsilon=0$ and analyzing the terms proportional to $\epsilon$ we find that $\alpha$ must satisfy the following equation\footnote{A subtlety in this calculation is that $\delta'(U) G_{ij,V} = - \delta(U)G_{ij,UV}$. At order $\epsilon^2$ the terms are proportional to $U^2 \delta(U)^2$ and this can be consistently taken as zero \cite{Sfetsos-94}.}
\be
\delta(U) \,G^{ij}\,\big(A\, \partial_i \partial_j -\frac{1}{2}G_{ij, UV}\,  \big)\, \alpha(t,x^i) = 8\, \pi\,G_\mt{N}\,T^\mt{shock}_{UU}\,,
\ee
while the metric functions and the energy-momentum tensor of the matter fields must be such that
\be
A_{,V} = G_{ij,V}=T^\mt{matter}_{VV}=0 \,\,\,\, \textrm{at}\,\,U=0\,.
\label{cond-sw}
\ee
In terms of the coordinates $t$ and $u$ of (\ref{metric}), the equation for $\alpha$ can be written as
\be
G^{ij}(\uh)\,\big[A(\uh)\, \partial_i \partial_j -\frac{\uh}{2}G_{ij}'(\uh)\,  \big]\, \alpha(t,x^i) = 8\, \pi\,G_\mt{N}\, E\, e^{2 \pi t / \beta}\, a(x^i)\,.
\label{eq-alpha}
\ee
This equation can be easily solved in the case of a homogeneous perturbation ($a=$ constant) by assuming $\alpha =$ constant $\times \, e^{2 \pi t / \beta}$. In the case of localized perturbations we solve (\ref{eq-alpha}) for two different situations. We first consider the case in which $a(z)=\delta(z)$ and $\alpha=\alpha(t,z)$. That means the perturbation propagates only in the anisotropic direction. Then we consider the case in which $a(x)=\delta(x)$ and the perturbation propagates in the $x$-direction, $\alpha = \alpha(t,x)$. In the first case we can rewrite the equation for $\alpha(t,z)$ as
\be
\big(\partial_z^2-M_{||}^2\big)\alpha(t,z)=\frac{G_{zz}(\uh)}{A(\uh)}8\, \pi\,E\, e^{2 \pi t / \beta}\, \delta(z)\,,
\label{alphaPara}
\ee
where
\be
M_{||}^2 =\frac{\uh\,G_{zz}}{2\,A}\Big(2\,\frac{G_{xx}'}{G_{xx}} +\frac{G_{zz}'}{G_{zz}}\Big)\Big|_{u=\uh}\,.
\ee
The above equation can be even more simplified if one uses the near-horizon expression for $G_{tt}$ (see Eq. (\ref{GnearHorizon})) and writes $A(\uh) = 2\,\uh\,c_1$. For large $|z|$, the solution of Eq. (\ref{alphaPara}) has the form
\be
\alpha(t,z) \sim \text{exp}\Big[\,\frac{2\pi}{\beta}(t-t_*)-M_{||}\, z \Big]\,.
\ee
By comparing the above solution with the general form of $C(q,\vec{x})$ (see Eq. (\ref{eq-C})) one can see that the Lyapunov exponent saturates the chaos bound $\lambda_L=2\pi /\beta$, while the butterfly velocity along the anisotropic direction is given by
\be
v_\mt{B}^{||\,2}=\left( \frac{2\pi}{\beta M_{||}} \right)^2= \frac{G_{tt}'}{G_{zz}\left(2\frac{G_{xx}'}{G_{xx}}+\frac{G_{zz}'}{G_{zz}} \right)}\Big|_{u=\uh}\,.
\label{VBpara}
 \ee
In the second case the equation for $\alpha(t,x)$ reads
\be
\big(\partial_x^2-M_{\perp}^2\big)\alpha(t,x)=\frac{G_{xx}(\uh)}{A(\uh)}8\, \pi\,E\, e^{2 \pi t / \beta}\, \delta(x)\,,
\label{alphaPerp}
\ee
where
\be
M_{\perp}^2 =\frac{\uh\,G_{xx}}{2\,A}\Big(2\,\frac{G_{xx}'}{G_{xx}} +\frac{G_{zz}'}{G_{zz}}\Big)\Big|_{u=\uh}\,.
\ee
The solution of (\ref{alphaPerp}) for large $|x|$ is given by
\be
\alpha(t,x) \sim \text{exp}\Big[\,\frac{2\pi}{\beta}(t-t_*)-M_{\perp}\, x \Big]\,.
\ee
This asymptotic behaviour implies, again, a maximal Lyapunov exponent $\lambda_L=2\pi/\beta$. The butterfly velocity orthogonal to the anisotropic direction is given by
\be
v_\mt{B}^{\perp\,2}=\left( \frac{2\pi}{\beta M_{\perp}} \right)^2 =\frac{G_{tt}'}{G_{xx}\left(2\frac{G_{xx}'}{G_{xx}}+\frac{G_{zz}'}{G_{zz}} \right)}\Big|_{u=\uh} \,.
\label{VBperp}
 \ee
The scrambling time in both cases can be estimated as
\be
t_*=\frac{\beta}{2\pi}\,\text{log}\,\Big( \frac{A(\uh)}{8\,\pi\,G_{ii}(\uh) G_\mt{N}}\Big) \approx \frac{\beta}{2\pi}\,\text{log}\,S_{\mt{BH}}\,,
\ee
where $G_{ii}$ can be $G_{zz}$ or $G_{xx}$, and $S_{\mt{BH}}$ is the Bekenstein-Hawking entropy.

\begin{figure}[H]
\begin{center}
\setlength{\unitlength}{1cm}
\includegraphics[width=0.6\linewidth]{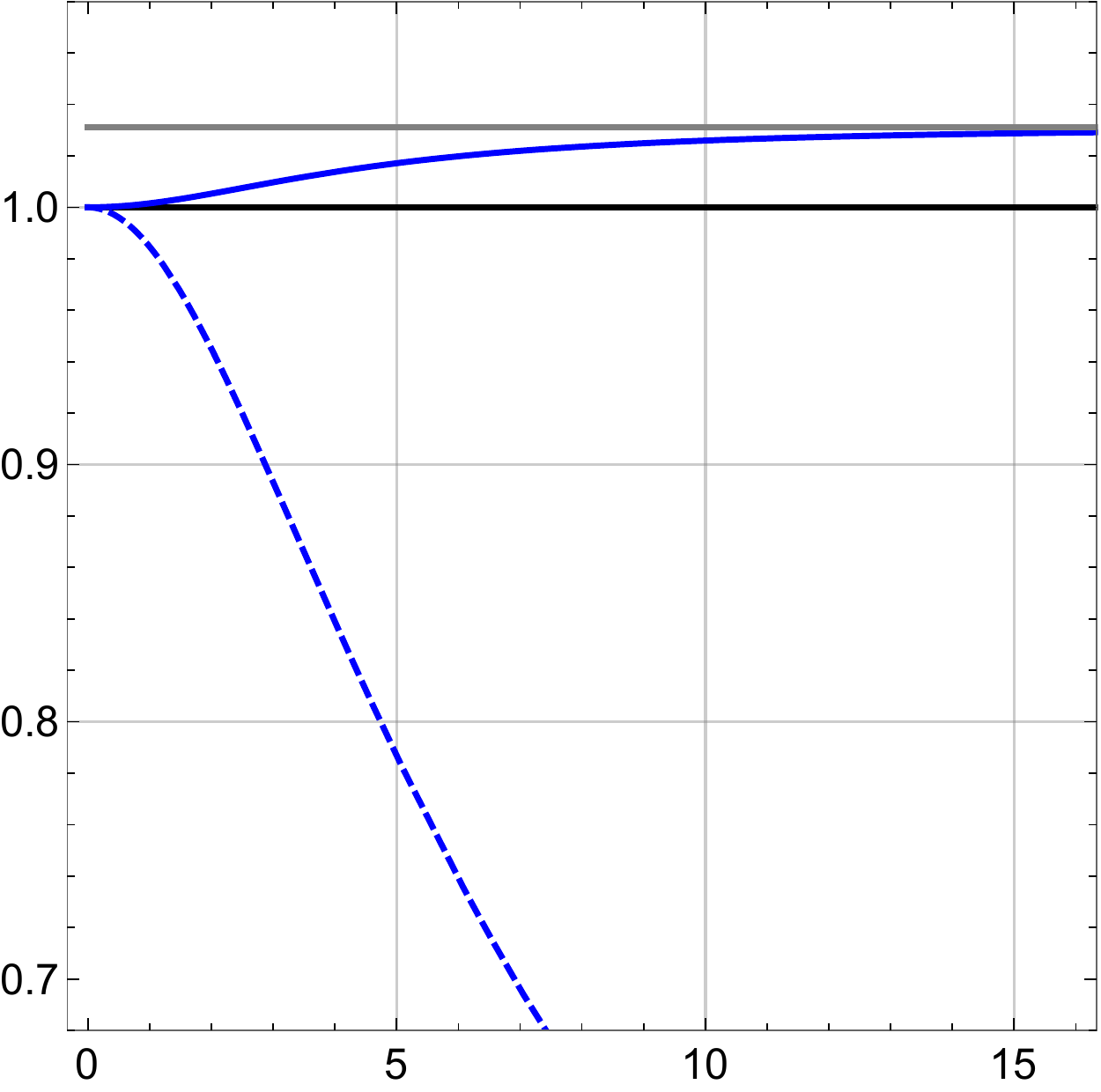} 
\put(-4.3,-.5){\large $a/T$}
\put(-10.,+5.1){\large $\frac{3}{2}v_\mt{B}^2$}
\put(-2.7,+7.5){\scriptsize $\frac{3}{2}v_\mt{B}^{\perp\,2}$}
\put(-7.,+5.6){\scriptsize $\frac{3}{2}v_\mt{B}^{||\,2}$}
\put(-7.0,+6.9){\scriptsize $\frac{3}{2}v_\mt{B}^{\mt{iso}\,2}$}
\put(-7.0,+8.2){\scriptsize $\frac{3}{2}v_\mt{B}^{\mt{Lif}\,2}$}

\end{center}
\caption{ \small
Butterfly velocity as a function of $a/T$ for the MT model. The continuous blue curve represent the result for $\frac{3}{2}v_\mt{B}^{\perp\,2}$, while the dashed blue curve represents the result for $\frac{3}{2}v_\mt{B}^{||\,2}$. The black horizontal line is the isotropic result $\frac{3}{2}v_\mt{B}^{\mt{iso}\,2}=1$, whereas the gray horizontal line is the result for a five-dimensional Lifshitz-like geometry, $\frac{3}{2}v_\mt{B}^{\mt{Lif}\,2}=\frac{33}{32}$.}
\label{fig-vB}
\end{figure}
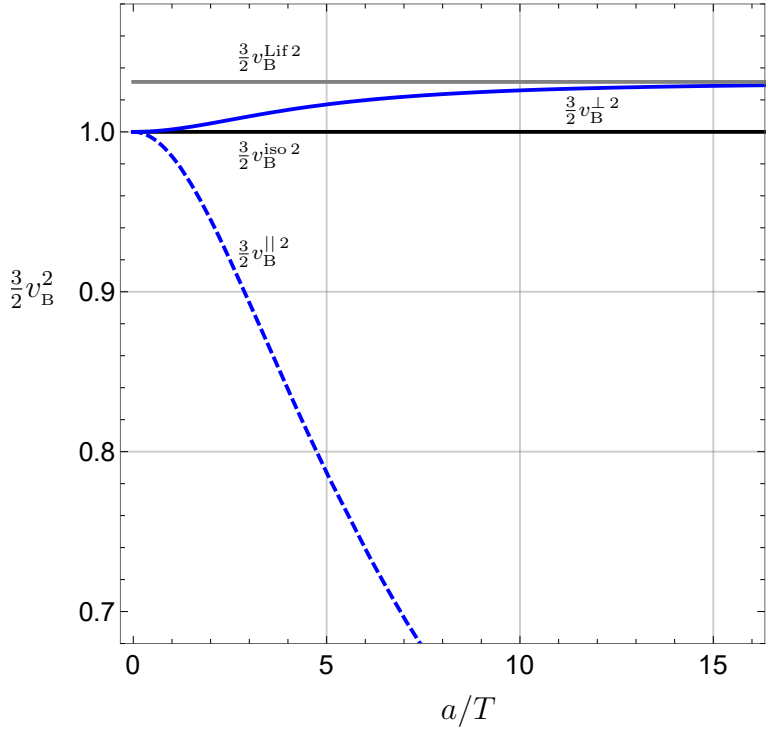

In figure \ref{fig-vB} we specialize our butterfly velocity formulas for the MT model. This is a black brane solution of type IIB supergravity that is spatially anisotropic. The effects of the anisotropy on the geometry are controlled by the ratio $a/T$, where $a$ is the parameter of anisotropy and $T$ is the black brane Hawking temperature. This solution describes a renormalization group flow from an AdS geometry in the ultraviolet, when $a/T$ is small, to a Lifshitz-like geometry in the infrared, when $a/T$ is large. More details about the MT model are provided in appendix \ref{appC}. 


\section{Mutual Information} \label{sec-3}

In this section we study the mutual information $I(A,B)$ between identical regions $A$ and $B$ on the left and right boundary, respectively. For simplicity, we only consider the case of strip-like regions. As we are dealing with anisotropic systems, we consider two types of regions $0 < z < \ell$ (strip oriented orthogonally to the direction of anisotropy), and $0 < x < \ell$ (strip oriented along the anisotropic direction). We first study the mutual information as a function of the strip's width $\ell$ in the unperturbed geometry. From this analysis we can characterize the critical width $\ell_c$ below which the mutual information vanishes. We then study how the mutual information is disrupted in shock wave geometries. We show that the entanglement velocity plays an important role in this phenomenon. 

\subsection{Mutual information versus strip's width}

\subsubsection*{Region $0<z< \ell$}

This region is delimited by two hyperplanes $z=0$ and $z=\ell$. The appropriate embedding for the corresponding extremal surface is $X^m = (0,x,y,z(u),u)$. The induced metric on this surface is generically given by
\be
g_{a b} = \frac{\partial X^m}{\partial \sigma^a} \frac{\partial X^n}{\partial \sigma^b} G_{m n}\,,
\ee
where $X^m,X^n$ and $G_{mn}$ denote the coordinates and the metric components of the five-dimensional geometry and $\sigma^a, \sigma^b$ and $g_{a b}$ denote the coordinates and the induced metric components on the extremal surface. In the above embedding $\sigma^{a}=(x,y,u)$. The components of the induced metric are given by
\bea
g_{xx} = g_{yy}=G_{xx}(u)\,,\qquad 
g_{uu} = G_{uu}(u)+z'(u)^2 G_{zz}(u)\,.
\eea
The area functional to be extremized is given by
\be
A^{\perp}=\int dx\, dy\, du\, \sqrt{g} = V_2 \int du\,|G_{xx}| \sqrt{G_{uu}+G_{zz} z'(u)^2}\,,
\label{eq-Aperp}
\ee
where $V_2 = \int dx dy$ denotes the volume of the hyperplanes delimiting the region $0<z<\ell$. As this functional does not depend explicitly on $z(u)$, there is a conserved quantity associated to translations in $z$ that is given by
\be
\gamma= \frac{\partial}{\partial z'} \big(|G_{xx}| \sqrt{G_{uu}+G_{zz} z'(u)^2} \big)=\frac{G_{xx} G_{zz} z'(u)}{\sqrt{G_{uu}+G_{zz} z'(u)^2}}=G_{xx}(u_m)\sqrt{G_{zz}(u_m)}\,,
\label{eq-gamma}
\ee
where the last equatily was obtained evaluating $\gamma$ at the turning point $u_m$ where $z'\rightarrow \infty$. Solving (\ref{eq-gamma}) for $z'$ one obtains
\be
G_{zz} z'^2=\frac{G_{uu}}{\gamma^{-2} G_{xx}^2 G_{zz}-1}\,.
\ee
Substituting the above result back in (\ref{eq-Aperp}) gives the extremal area
\be
A^{\perp}_{\text{ext}}= 2 V_2 \int_{0}^{u_m} du\,|G_{xx}| \sqrt{G_{uu}}\frac{1}{\sqrt{1-\gamma^2 G_{xx}^{-2}G_{zz}^{-1}}}\,.
\ee
The entanglement entropy for the regions $A$ and $B$ can then be obtained as
\be
S(A)=S(B)=\frac{A^{\perp}_{\text{ext}}}{4 G_\mt{N}}=\frac{V_2}{2 G_\mt{N}} \int_{0}^{u_m} du\,|G_{xx}| \sqrt{G_{uu}}\frac{1}{\sqrt{1-\gamma^2 G_{xx}^{-2}G_{zz}^{-1}}}\,.
\ee
To compute $S(A \cup B)$ we need to compute the area of the surface that passes through the horizon connecting both sides. There are two such surfaces, one corresponding to the hyperplane $z=0$ and other corresponding to the hyperplane $z=\ell$. By symmetry, the total area of these surfaces will be four times the area of a surface that extends from the boundary to the horizon, being given by
\be
4 V_2  \int_{0}^{u_\mt{H}} du\, |G_{xx}| \sqrt{G_{uu}}
\label{eq-areaAB}
\ee
and $S(A \cup B)$ can be calculated by dividing the above result by $4 G_\mt{N}$. The mutual information is then given by
\be
I_{\perp}(u_m)=\frac{V_2}{G_\mt{N}} \left[\int_{0}^{u_m} du\,|G_{xx}| \sqrt{G_{uu}}\frac{1}{\sqrt{1-\gamma^2 G_{xx}^{-2}G_{zz}^{-1}}} -  \int_{0}^{u_\mt{H}} du\,|G_{xx}| \sqrt{G_{uu}} \right]\,.
\label{eq-MIperp}
\ee
In order to study the mutual information as a function of strip's width, we write $\ell$ as a parametric function of $u_m$
\be
\ell_{\perp}(u_m)=\int dz = 2 \int_{0}^{u_m} du\, z'(u) = 2 \int_{0}^{u_m} du\,\sqrt{\frac{G_{uu}}{G_{zz}}} \frac{1}{\sqrt{\gamma^{-2}G_{xx}^2 G_{zz}-1}}\,.
\ee 
Using the above formulas one can plot $I_{\perp}$ as a function of $\ell_{\perp}$. We use the subscript $\perp$ in $\ell$ to indicate that this quantity has been calculated for a strip orthogonal to the anisotropic direction. 

\subsubsection*{Region $0<x<\ell$} 
In this case the appropriate embedding is $X^m = (0,x(u),y,z,u)$, the coordinates along the surface are $\sigma^a = (y,z,u)$. The components of the induced metric are
\bea
g_{yy} =G_{yy}\,, \qquad g_{zz} =G_{zz}\,,\qquad
g_{uu} =G_{uu}+G_{xx} x'(u)^2
\eea
and the functional to be extremized is
\be
A^{||}= \int dy\, dz\, du \sqrt{g} = V_2 \int du \sqrt{G_{xx} G_{zz}} \sqrt{G_{uu}+G_{xx} x'(u)^2}\,,
\ee
where $V_2 = \int dy \, dz$ is the volume of the hyperplanes $x=0$ and $x=\ell$. Proceeding as before we can show that the mutual information and the length $\ell$ are given by
\be
I_{||}(u_m)=\frac{V_2}{G_\mt{N}} \left[\int_{0}^{u_m} du\,\sqrt{G_{xx} G_{zz} G_{uu}}\frac{1}{\sqrt{1-\gamma^2 G_{xx}^{-2}G_{zz}^{-1}}} -  \int_{0}^{u_\mt{H}} du\, \sqrt{G_{xx} G_{zz} G_{uu}} \right]
\ee
and
\be
\ell_{||}(u_m)=\int dz = 2 \int_{0}^{u_m} du\, x'(u) = 2 \int_{0}^{u_m} du\,\sqrt{\frac{G_{uu}}{G_{xx}}} \frac{1}{\sqrt{\gamma^{-2}G_{xx}^2 G_{zz}-1}}\,,
\ee
where $\gamma$ = $G_{xx}(u_m) \sqrt{G_{zz}(u_m)}$.

\begin{figure}[H]
\begin{center}
\setlength{\unitlength}{1cm}
\includegraphics[width=0.6\linewidth]{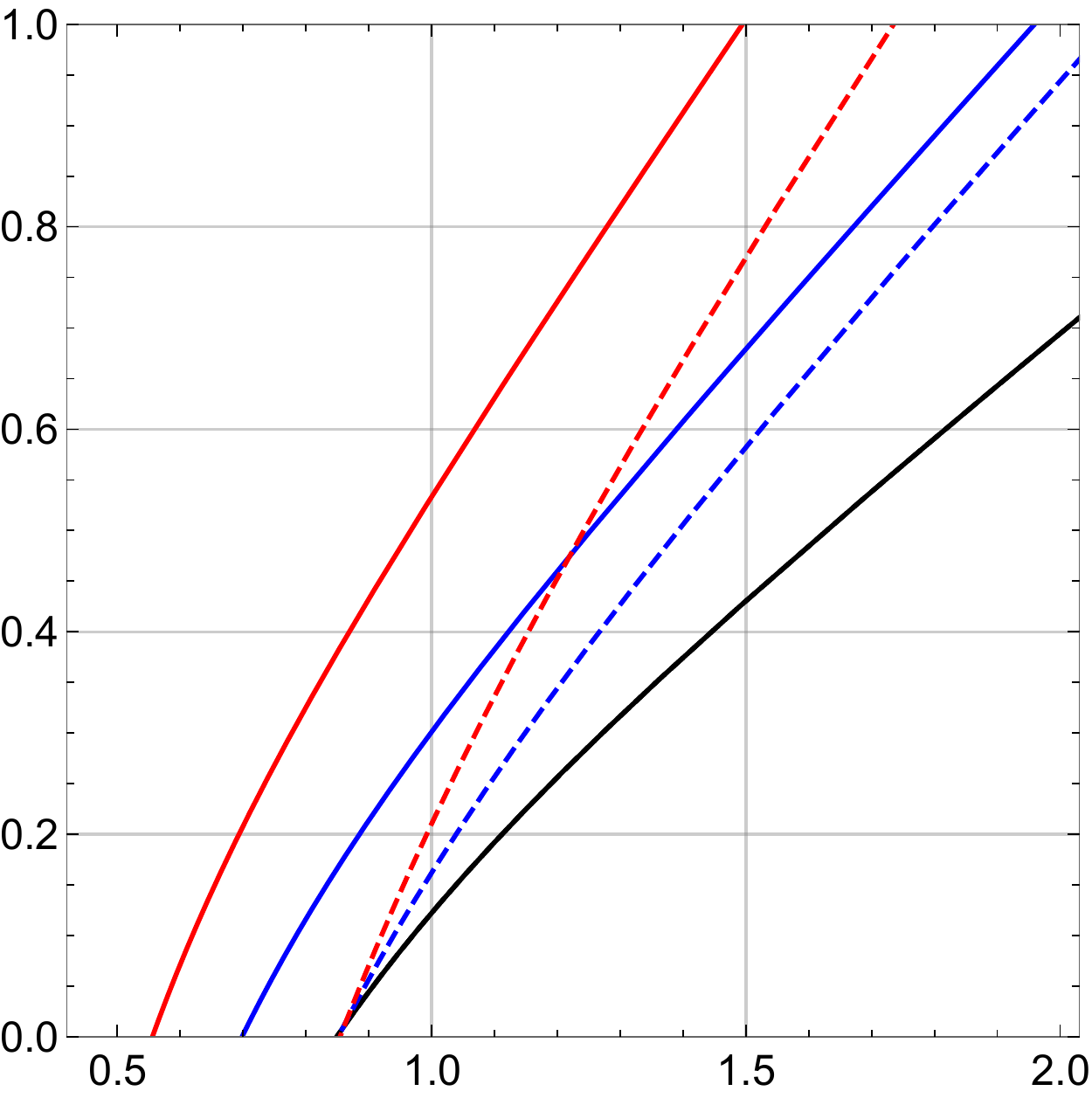} 
\put(-4.0,-.5){\large $\ell$}
\put(-10.,+4.7){\large $I(\ell)$}

\end{center}
\caption{ \small
Mutual Information (in units of $V_2/G_\mt{N}$) as a function of $\ell$ for the MT model. The curves correspond from the right to the left to $a/T = 0$ (black curve), $a/T = 8.56$ (blue curves) and $a/T = 21.57$ (red curves). The continuous/dashed curves represent the result for a strip orthogonal/parallel to the anisotropic direction. Here we have fixed $T=1/\pi$.}
\label{fig-MIversusL}
\end{figure}

In figure \ref{fig-MIversusL} we specialize the above formulas for the MT model. This figure shows the mutual information as a function of the strip's width $\ell$ for some values of the anisotropy parameter and for strips orthogonal and parallel to the anisotropic direction.

\subsection{Disruption of the mutual information}

In this section we study how the mutual information $I(A,B)$ is affected by a shock wave produced by a homogeneous perturbation. In this case the shift in the $V$-coordinate is simply given by $\alpha = \text{constant} \times e^{2\pi t_0/\beta}$. This parameter controls the strength of the shock wave. In this geometry the wormhole becomes longer, but the left and right exterior regions are unchanged. As a result, only probes that extend through the wormhole can diagnose the effects of the shock wave. This implies that the entanglement entropies $S(A)$ and $S(B)$ are not affected by the shock wave, because the corresponding surfaces can never penetrate the horizon \cite{Hubeny-2012}. The only piece of the mutual information that changes in the shock wave geometry is $S(A \cup B)$, since the corresponding extremal surface extends through the wormhole. In fact, as we move the time $t_0$ at which the perturbation was applied further into the past, the wormhole becomes longer, generating an increase of $S(A \cup B)$ and a corresponding decrease of $I(A,B)$. Therefore, for an early enough perturbation, $t_0 \gtrsim t_*$, the mutual information drops to zero, signalizing the complete disruption of the local pattern of entanglement of the geometry. 

In the computation of $S(A \cup B)$ we follow \cite{Leichenauer-2014} and consider the case where $A=B$ is half the space. This simplifies the analysis because the corresponding extremal surface divides the transverse space $(x,y,z)$ in half, and the minimization problem is reduced to a two-dimensional problem. 
To account for the anisotropic background, we consider two types of regions $(0<z<\infty)$ and $(0<x<\infty)$. As $S(A \cup B)$ is independent of the strip's width $\ell$, we can compute it for $\ell \rightarrow \infty$ and use the obtained result to compute $S(A \cup B)$ for a finite $\ell$, as done in \cite{Sircar-2016}.\footnote{In this case we should multiply the result by two to account for the two extremal surfaces, one at $z=0$ and the other one at $z=\ell$ (or $x=0$ and $x=\ell$).} 

\subsubsection*{Region $0 < z < \infty$} 
\begin{figure}[h!]
\centering

\begin{tikzpicture}[scale=1.5]

\draw [thick,decorate,decoration={zigzag,segment length=2mm, amplitude=0.5mm}]  (5.5,3) -- (9.5,3);

\draw [thick,decorate,decoration={zigzag,segment length=2mm, amplitude=0.5mm}]  (6.6,0) -- (10.6,0);

\draw [thick] (6.5,0) -- (9.5,3);
\draw [thick] (6.6,0) -- (9.6,3);

\draw [thick] (5.5,3) -- (6.5,0);

\draw [thick] (9.6,3) -- (10.6,0);

\draw [thick,dashed] (5.5,3) -- (7.5,1);
\draw [thick,dashed] (8.6,2) -- (10.6,0);

\draw[very thick, red] (6,1.5) -- (10.1,1.5);

\draw[thick,red, |-|] (7,1.5) -- (7.5,1.5);

\draw[thick,red, |-|] (7,1.5) -- (8,1.5);

\draw [thick,<->] (8.13,1.4) -- (7.63,.9);

\node [scale=.8] at (8.,1.0) {$\frac{\alpha}{2}$};

\draw [dashed, blue] (5.5,3) to (7.1,1.65) to [out=-35,in=180] (7.5,1.52)
to [out=0,in=-145] (7.9,1.65) to (9.5,3);

\node [scale=.7] at (7.5,1.7) {$u_0$};

\node [scale=.45] at (6.5,1.39) {$I$};
\node [scale=.45] at (7.27,1.39) {$II$};
\node [scale=.45] at (7.73,1.39) {$III$};



\end{tikzpicture}
\vspace{0.1cm}
\caption{ \small Extremal surface (horizontal, red) in the shock wave geometry. Following \cite{Leichenauer-2014}, we divide the left half of the surface into three parts, $I$, $II$ and $III$. The segments $II$ and $III$ have the same area and they are separated by the point $u_0$ at which the surface defined by $u=u_0$ (blue, dashed curve) intersects the extremal surface.}
\label{fig-surfaceLocation}
\end{figure}
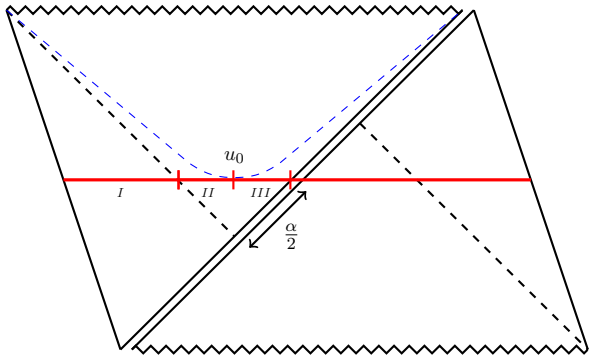

The appropriate embedding in this case is $X^m = (t,x,y,0,u(t))$. The components of the induced metric are
\bea
g_{xx} = g_{yy}=G_{xx}(u)\,,\qquad 
g_{tt} =-G_{tt}+G_{uu}\dot{u}^2\,.
\eea
The functional to be extremized is then
\be
A^{\perp} = \int dx dy dt \sqrt{g} = V_2 \int dt \,|G_{xx}| \sqrt{-G_{tt}+G_{uu}\dot{u}^2} = V_2 \int dt\, \mathcal{L}(u,\dot{u})\,.
\label{eq-AreaPerpAlpha}
\ee
This functional is invariant under translations in $t$ and the associated conserved quantity is given by
\be
\gamma_{\perp}=\frac{\partial \mathcal{L}}{\partial \dot{u}}\dot{u}-\mathcal{L}= \frac{|G_{xx}|G_{tt}}{\sqrt{-G_{tt}+G_{uu}\dot{u}^2}}=-|G_{xx}(u_0)|\sqrt{-G_{tt}(u_0)}\,,
\label{eq-gammaPerp}
\ee
where in the last equality we compute $\gamma_{\perp}$ in the point $u_0$ at which $\dot{u}=0$ (this point lies behind the outer horizon, see figure \ref{fig-surfaceLocation}). The limit at which the shockwave is absent $\alpha \rightarrow 0$ is reached when $u_0 \rightarrow \uh$, because in this case $\gamma_{\perp} \rightarrow 0$ and we recover the result of Eq. (\ref{eq-areaAB}).

Solving Eq. (\ref{eq-gammaPerp}) for $\dot{u}$ and substituting in Eq. (\ref{eq-AreaPerpAlpha}) we find
\be
A^{\perp}_{\text{ext}}(u_0) = 2\,V_2 \int du\,|G_{xx}| \sqrt{G_{uu}} \frac{1}{\sqrt{1+\gamma_{\perp}^2 G_{xx}^{-2}G_{tt}^{-1}}}\,.
\label{Aext-u0}
\ee
It is convenient to split the integral in Eq. (\ref{Aext-u0}) into three regions, $I$, $II$ and $III$. See figure \ref{fig-surfaceLocation}. As the regions $II$ and $III$ have the same area, we can write $\int_{I \cup II \cup III} = \int_{0}^{\uh}+ 2 \int_{\uh}^{u_0}$. To use the above result to compute the mutual information for strips of finite width, one should multiply the result of Eq. (\ref{Aext-u0}) by two to account for the two extremal surfaces bounding the strip.

The entanglement entropy of the region $A \cup B$ can then be obtained as
\be
S(A \cup B) =\frac{A^{\perp}_{\text{ext}}(u_0)}{4G_\mt{N}} = \frac{V_2}{2G_\mt{N}} \int du\,|G_{xx}| \sqrt{G_{uu}} \frac{1}{\sqrt{1+\gamma_{\perp}^2 G_{xx}^{-2}G_{tt}^{-1}}}\,.
\label{Sperp-u0}
\ee
The above equation shows that $S(A \cup B)$ is a function of $u_0$. To understand how this behaviour is related to the shock wave parameter $\alpha$ we need to find a relation between $\alpha$ and $u_0$. We show in appendix \ref{appA} that the relation between $\alpha$ and $u_0$ is given by
\be
\alpha(u_0)_{\perp}=2\, e^{K_1^{\perp}(u_0)+K_2^{\perp}(u_0)+K_3^{\perp}(u_0)}\,,
\label{alphaPerp-u0}
\ee
where
\bea
&K_1^{\perp}&=-\frac{4 \pi}{\beta}  \int_{\bar{u}}^{u_0} du \sqrt{\frac{G_{uu}}{G_{tt}}}\,,\cr
&K_2^{\perp}&= \frac{2 \pi}{\beta} \int_{0}^{u_\mt{H}} du \sqrt{\frac{G_{uu}}{G_{tt}}} \left(1-\frac{1}{\sqrt{1+G_{xx}^2 G_{tt} \gamma_{\perp}^{-2}}} \right)\,,\cr
&K_3^{\perp}&=-\frac{4 \pi}{\beta} \int_{u_\mt{H}}^{u_0} du \sqrt{\frac{G_{uu}}{G_{tt}}} \left(1-\frac{1}{\sqrt{1+G_{xx}^2 G_{tt} \gamma_{\perp}^{-2}}} \right)\,,
\eea
where $\bar{u}$ is a point behind the outer horizon at which $u_* = 0$. The limit in which the shock wave is absent is achieved when $u_0 = \uh$ because $\alpha_{\perp}(\uh)=0$. The function $\alpha_{\perp}(u_0)$ increases as we move $u_0$ deeper behind the horizon and diverges at some point $u_0=u_c^{\perp}$. See figure \ref{uc-alpha}. We show in appendix \ref{appB} that this point is given implicitly by the equation
\be
\frac{\big( G_{xx}^2 G_{tt}\big)'}{ G_{xx}^2 G_{tt}}\Big|_{u=u_c^{\perp}}=0\,.
\ee

Using Eq. (\ref{Sperp-u0}) and (\ref{alphaPerp-u0}) one can plot $S(A \cup B)$ as a function of the shock wave parameter $\alpha$. This function $S(A \cup B)(\alpha)$ has a $\alpha$-independent divergence that can be subtracted by considering the regularized quantity \cite{Sircar-2016}
\be
S^{\text{reg}}(A \cup B)(\alpha)=S(A \cup B)(\alpha)-S(A \cup B)(\alpha=0)\,.
\ee
With the above definition, the mutual information can be written as \cite{Sircar-2016}
\be
I(A,B;\alpha)=S(A)+S(B)-S(A \cup B)(\alpha)=I(\ell)-S^{\text{reg}}(A \cup B)(\alpha)\,,
\label{Eq-Ialpha}
\ee
where $I(\ell)=I(A,B;\alpha=0)$ is the mutual information calculated in the absence of the shock wave (given by Eq. (\ref{eq-MIperp})).

\subsubsection*{Region $0 < x < \infty$}
The appropriate embedding in this case is $X^m = (t,x,0,z,u(t))$. The components of the induced metric are
\bea
g_{xx} = G_{xx}\,, \qquad g_{zz} = G_{zz}\,,\qquad 
g_{tt} =-G_{tt}+G_{uu}\dot{u}^2\,.
\eea
The functional to be extremized is then
\be
A^{||} = \int dx dz dt \sqrt{g} = V_2 \int dt\, \sqrt{G_{xx} G_{zz}} \sqrt{-G_{tt}+G_{uu}\dot{u}^2} = V_2 \int dt \,\mathcal{L}(u,\dot{u})\,.
\label{eq-AreaParaAlpha}
\ee
Proceeding as before we can show that the extremal area is given by
\be
A^{||}_{\text{ext}}(u_0) = V_2 \int du\,\sqrt{G_{xx} G_{zz} G_{uu}} \frac{1}{\sqrt{1+\gamma_{||}^2 G_{xx}^{-1} G_{zz}^{-1}G_{tt}^{-1}}}\,,
\ee
where
\be
\gamma_{||}=-\sqrt{G_{xx}(u_0) G_{zz}(u_0)}\sqrt{-G_{tt}(u_0)}\,.
\label{eq-gammaPara}
\ee
The entanglement entropy of the region $A \cup B$ can then be computed as
\be
S(A \cup B) =\frac{A^{||}_{\text{ext}}(u_0)}{4G_\mt{N}} = \frac{V_2}{2G_\mt{N}} \int du\,\sqrt{G_{xx} G_{zz} G_{uu}} \frac{1}{\sqrt{1+\gamma_{||}^2 G_{xx}^{-1} G_{zz}^{-1}G_{tt}^{-1}}}\,.
\label{Spara-u0}
\ee
As shown in appendix \ref{appA}, the relation between the shock wave parameter $\alpha$ and $u_0$ is now given by 
\be
\alpha_{||}(u_0)=2\, e^{K_1^{||}(u_0)+K_2^{||}(u_0)+K_3^{||}(u_0)}\,,
\label{alphaPara-u0}
\ee
where
\bea
&K_1^{||}&=-\frac{4 \pi}{\beta} \int_{\bar{u}}^{u_0} du \sqrt{\frac{G_{uu}}{G_{tt}}}\,,\cr
&K_2^{||}&= \frac{2 \pi}{\beta} \int_{0}^{u_\mt{H}} du \sqrt{\frac{G_{uu}}{G_{tt}}} \left(1-\frac{1}{\sqrt{1+G_{xx} G_{zz} G_{tt} \gamma_{||}^{-2}}} \right)\,,\cr
&K_3^{||}&=-\frac{4 \pi}{\beta} \int_{u_\mt{H}}^{u_0} du \sqrt{\frac{G_{uu}}{G_{tt}}} \left(1-\frac{1}{\sqrt{1+G_{xx} G_{zz} G_{tt} \gamma_{||}^{-2}}} \right)\,,
\eea
where $\bar{u}$ is a point behind the outer horizon at which $u_* = 0$. The function $\alpha_{||}(u_0)$ is zero when $u_0=\uh$, corresponding to the absence of the shock wave, and increases as we move $u_0$ deeper behind the horizon, diverging at some point $u_0=u_c^{||}$. See figure \ref{uc-alpha}. We show in appendix \ref{appB} that this point is implicitly given by
\be
\frac{\big(G_{xx}G_{zz} G_{tt}\big)'}{ G_{xx}G_{zz} G_{tt}}\Big|_{u=u_c^{||}}=0\,.
\label{Eq-ucPara}
\ee

The behaviour of $S(A \cup B)$ as a function of $\alpha$ can be studied using Eq. (\ref{Spara-u0}) and (\ref{alphaPara-u0}). The regularized version of $S(A \cup B)$ can be defined as before.

\begin{figure}[H]
\begin{center}
\begin{tabular}{cc}
\setlength{\unitlength}{1cm}
\hspace{-0.9cm}
\includegraphics[width=6.4cm]{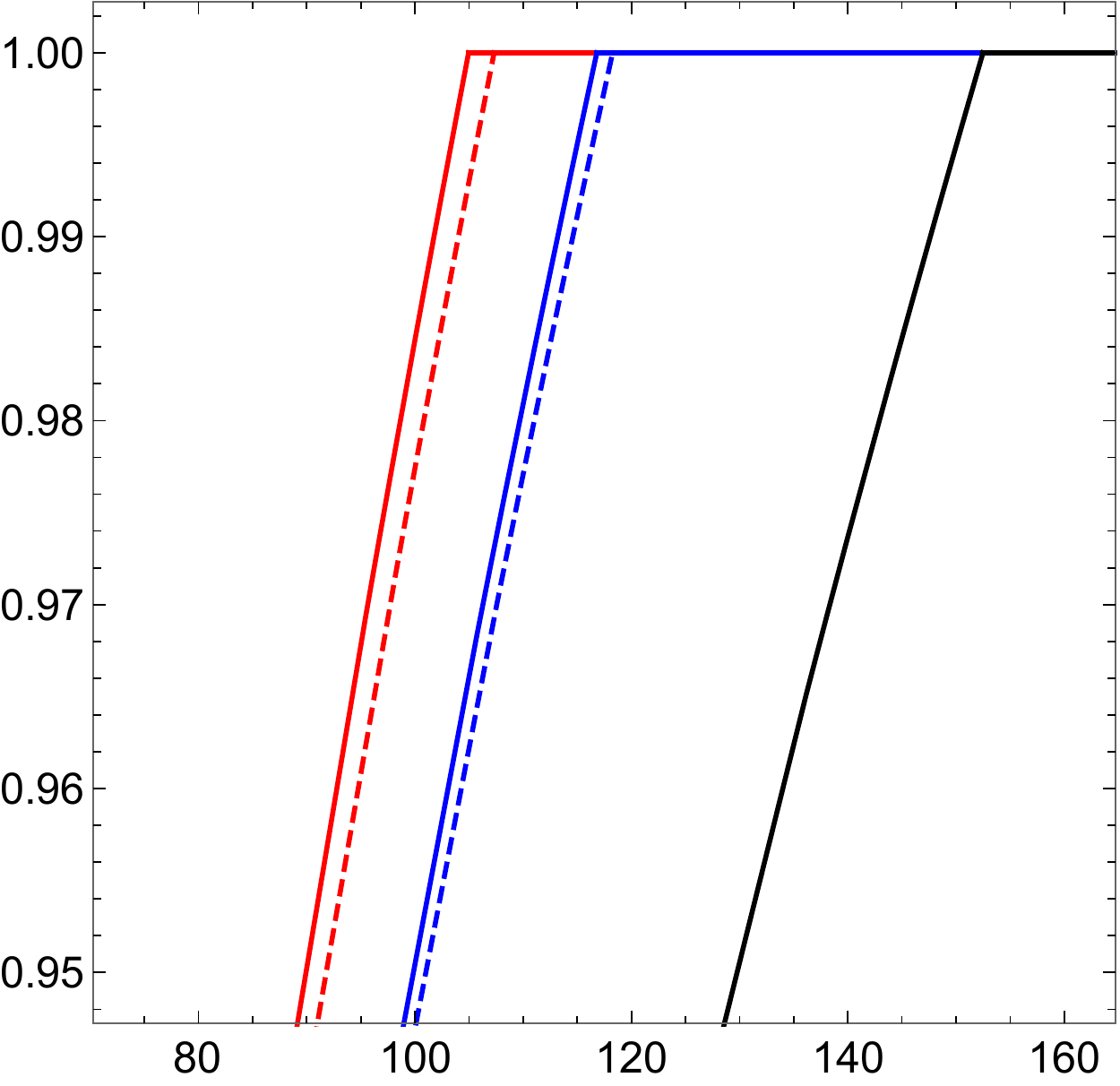} 
\qquad\qquad & 
\includegraphics[width=6.5cm]{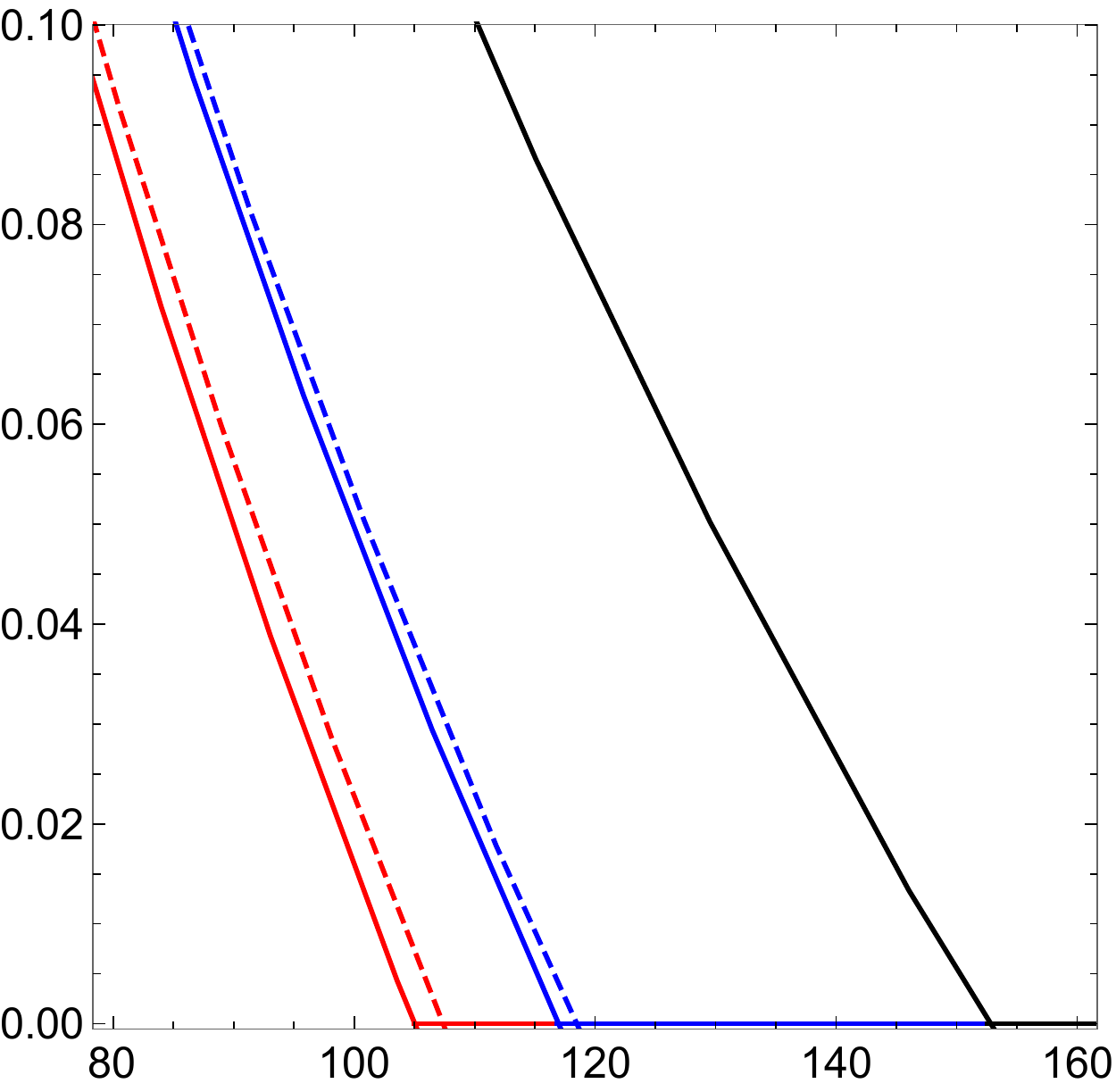}
\qquad
  \put(-420,60){\rotatebox{90}{\large $S^{\text{reg}}(A \cup B)$}}
         \put(-305,-10){\large $\alpha$}
         \put(-205,75){\rotatebox{90}{\large $I(A,B)$}}
         \put(-87,-10){\large $\alpha$}
         \put(-87,-30){$(b)$}
         \put(-315,-30){$(a)$}
         
\end{tabular}
\end{center}
\caption{ \small (a) $S^{\text{reg}}(A \cup B)$ as a function of $\alpha$. (b) Mutual information $I(A,B)$ (in units of $V_2/G_\mt{N}$) as a function of $\alpha$. All the curves have the same mutual information $I(\ell)=1$  at $\alpha =0$.  Both in (a) and (b) the curves correspond to $a/T = 0$ (black curve), $a/T = 8.56$ (blue curves) and $a/T = 21.57$ (red curves). The continuous/dashed curves represent the result for a strip orthogonal/parallel to the anisotropic direction. Here we have fixed $T=1/\pi$.}
\label{MI-SAB}
\end{figure}
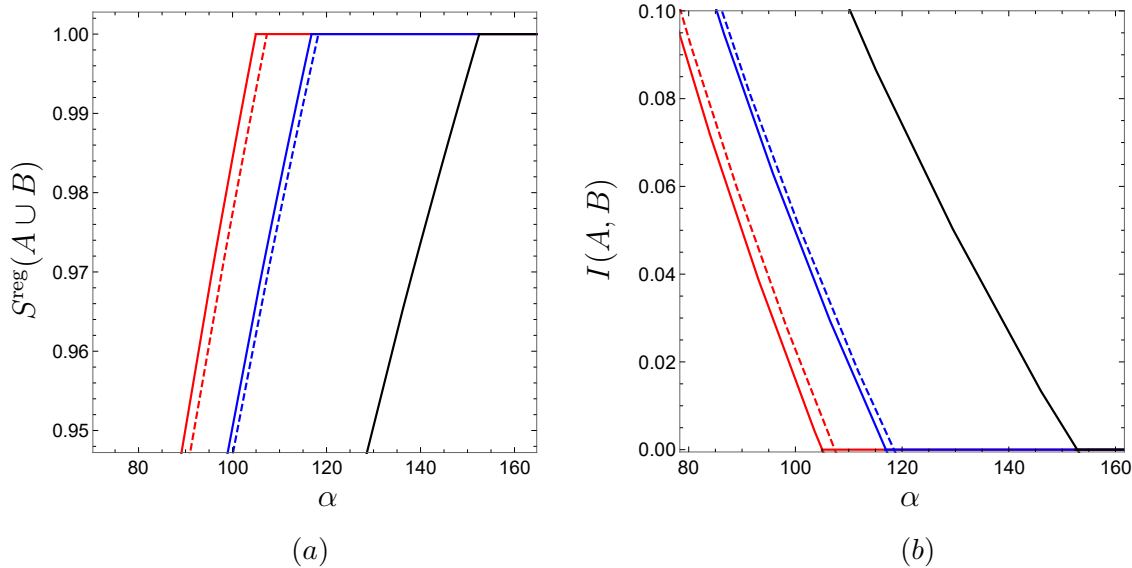

We now specialize some of the above results for the MT model. The figure \ref{uc-alpha} (a) shows how the critical points $u_c^{\perp}$ and $u_c^{||}$ change as we increase the anisotropy parameter. Figure  \ref{uc-alpha} (b) shows the behaviour of shock wave parameter $\alpha$ as a function of $u_0$ for several values of the anisotropy and for strips parallel and orthogonal to the anisotropic direction. In figure \ref{MI-SAB} we show how the regularized entanglement entropy $S^\text{reg}(A \cup B)$ and the mutual information $I(A,B)$ behave as a function of the shock wave parameter $\alpha$ for different strip orientations and for several values of the anisotropy parameter. For small values of the shock wave parameter $\alpha$ the entanglement entropy $S(A \cup B)$ is given by the area of the extremal surfaces connecting the two sides of the geometry. As these surfaces probe the interior of the black brane, they are affected by the shock wave, and they become larger as we increase $\alpha$. Both $S^\text{reg}(A \cup B)$ and $I(A,B)$ display a sharp transition to a constant value for a given value of the shock wave parameter $\alpha$. This happens when the area of the extremal surfaces connecting the two sides of the black brane geometry becomes larger than the area of the static U-shaped surfaces which are separately homologous to the regions $A$ and $B$. In this case the minimal area surfaces are the ones lying outside the horizon and the entanglement entropy of $A \cup B$ assumes the value $S(A \cup B)= S(A)+S(B)$, which does not depend on $\alpha$, because the corresponding extremal surfaces are not affected by the shock wave. In this situation the mutual information assumes a constant zero value, which implies that the regularized entanglement entropy is equal to the initial mutual information $S^\text{reg}(A \cup B) = I(\ell)$ (see Eq. (\ref{Eq-Ialpha})).

\begin{figure}[H]
\begin{center}
\begin{tabular}{cc}
\setlength{\unitlength}{1cm}
\hspace{-0.9cm}
\includegraphics[width=6.5cm]{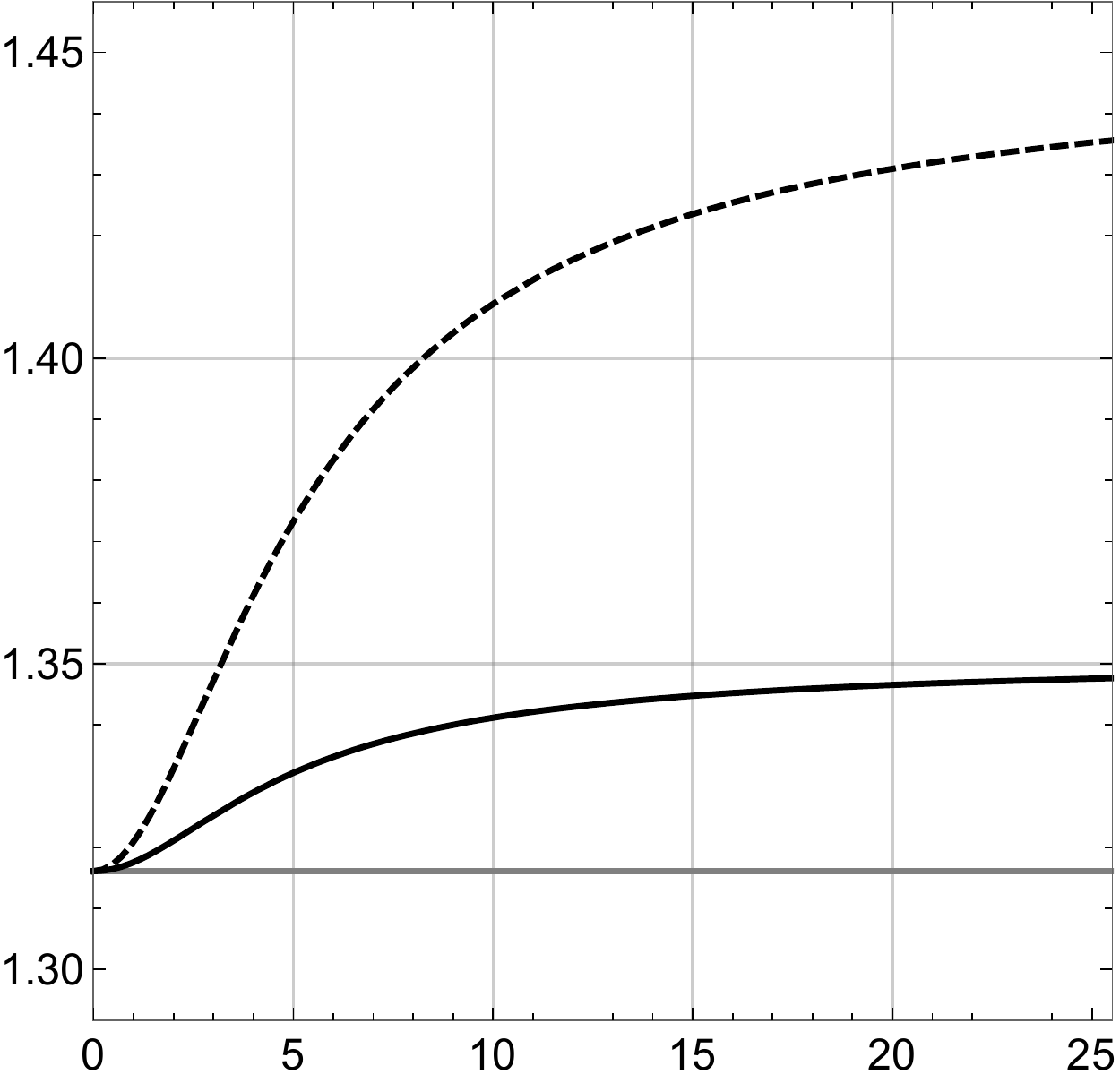} 
\qquad\qquad & 
\includegraphics[width=6.5cm]{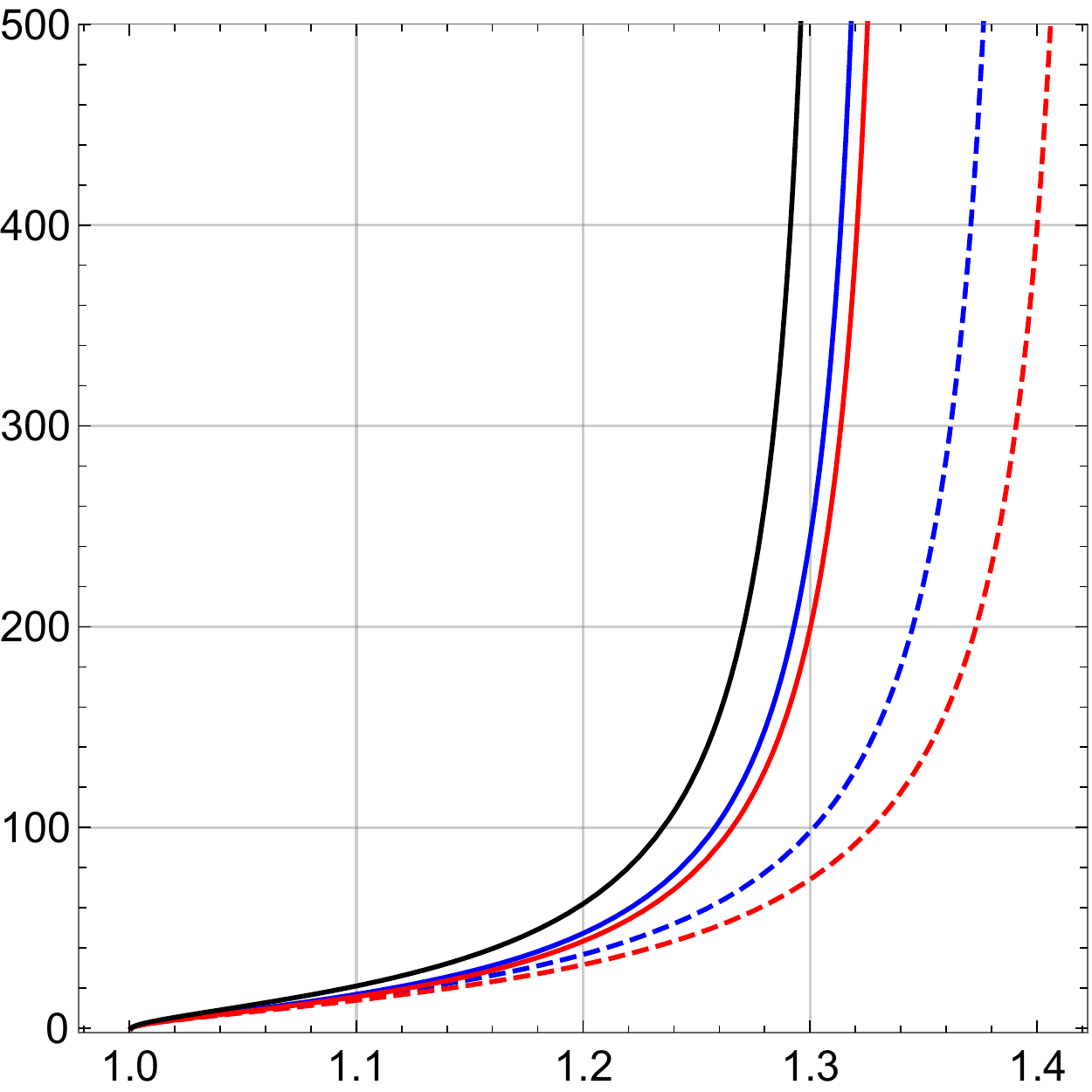}
\qquad
  \put(-435,100){\large $u_c/\uh$}
         \put(-310,-15){\large $a/T$}
         \put(-195,100){\large $\alpha$}
         \put(-87,-15){\large $u_0/\uh$}
         \put(-82,-33){$(b)$}
         \put(-305,-33){$(a)$}
         \put(-300,70){$u_c^{\perp}/\uh$}
         \put(-300,152){$u_c^{||}/\uh$}
         \put(-300,40){$u_c^{\mt{iso}}/\uh$}
\end{tabular}
\end{center}
\caption{ \small(a) Critical point $u_c$ at which $\alpha$ diverges as a function of the anisotropy parameter. The continuous curve represent the result for $u_c^{\perp}$, while the dashed curve represent the result for $u_c^{||}$. The horizontal gray curve is the result for the isotropic case $u_c^{\mt{iso}}/\uh =3^{1/4}$. (b) The shock wave parameter $\alpha$ as a function of $u_0$. The curves correspond from the left to the right to $a/T = 0$ (black curve), $a/T = 8.56$ (blue curves) and $a/T = 21.57$ (red curves). The continuous/dashed curves represent the result for a strip orthogonal/parallel to the anisotropic direction.}
\label{uc-alpha}
\end{figure}

\subsection{Spreading of entanglement}
\begin{figure}[H]
\begin{center}
\begin{tabular}{cc}
\setlength{\unitlength}{1cm}
\hspace{-0.9cm}
\includegraphics[width=6.3cm]{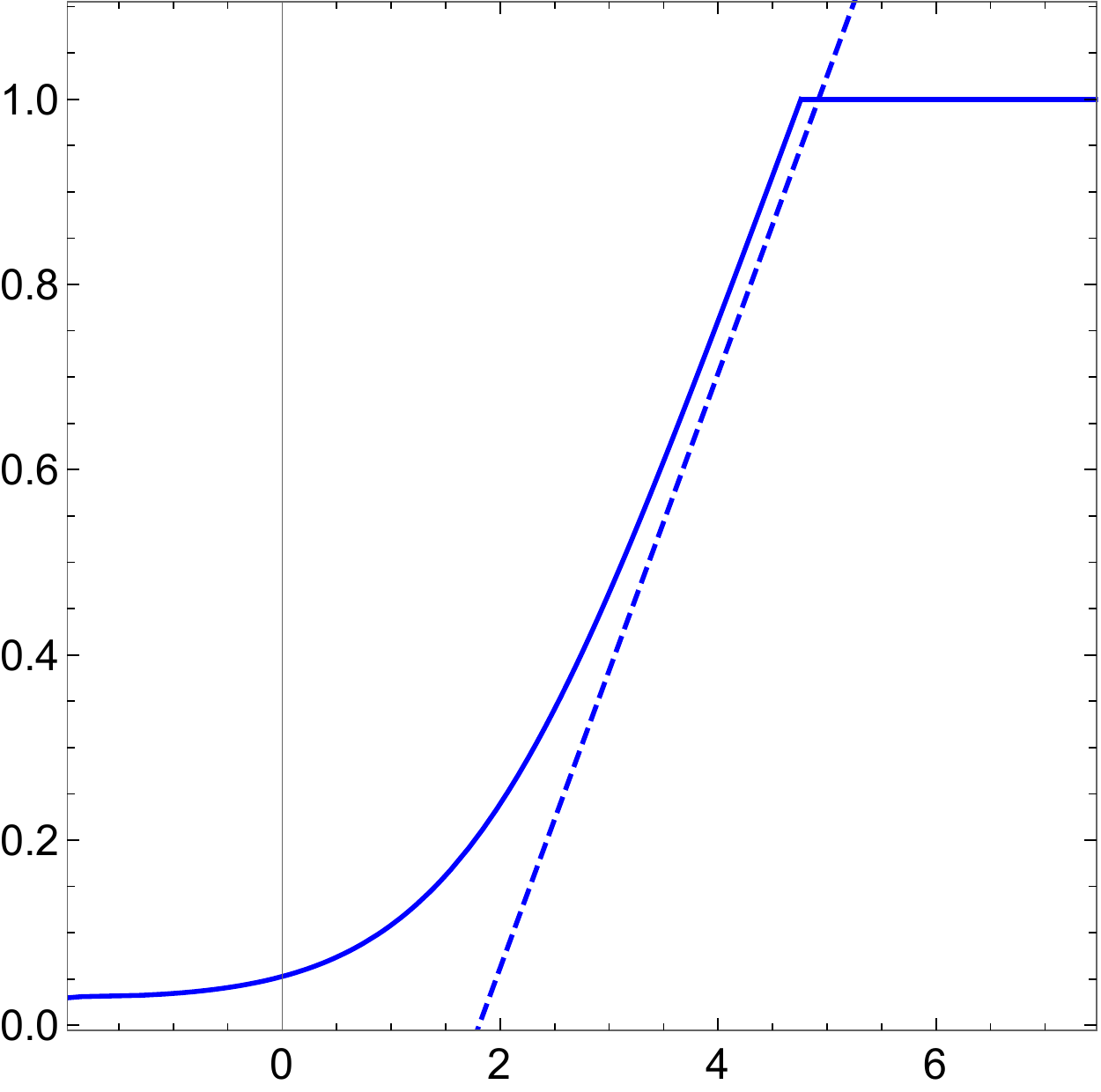} 
\qquad\qquad & 
\includegraphics[width=6.7cm]{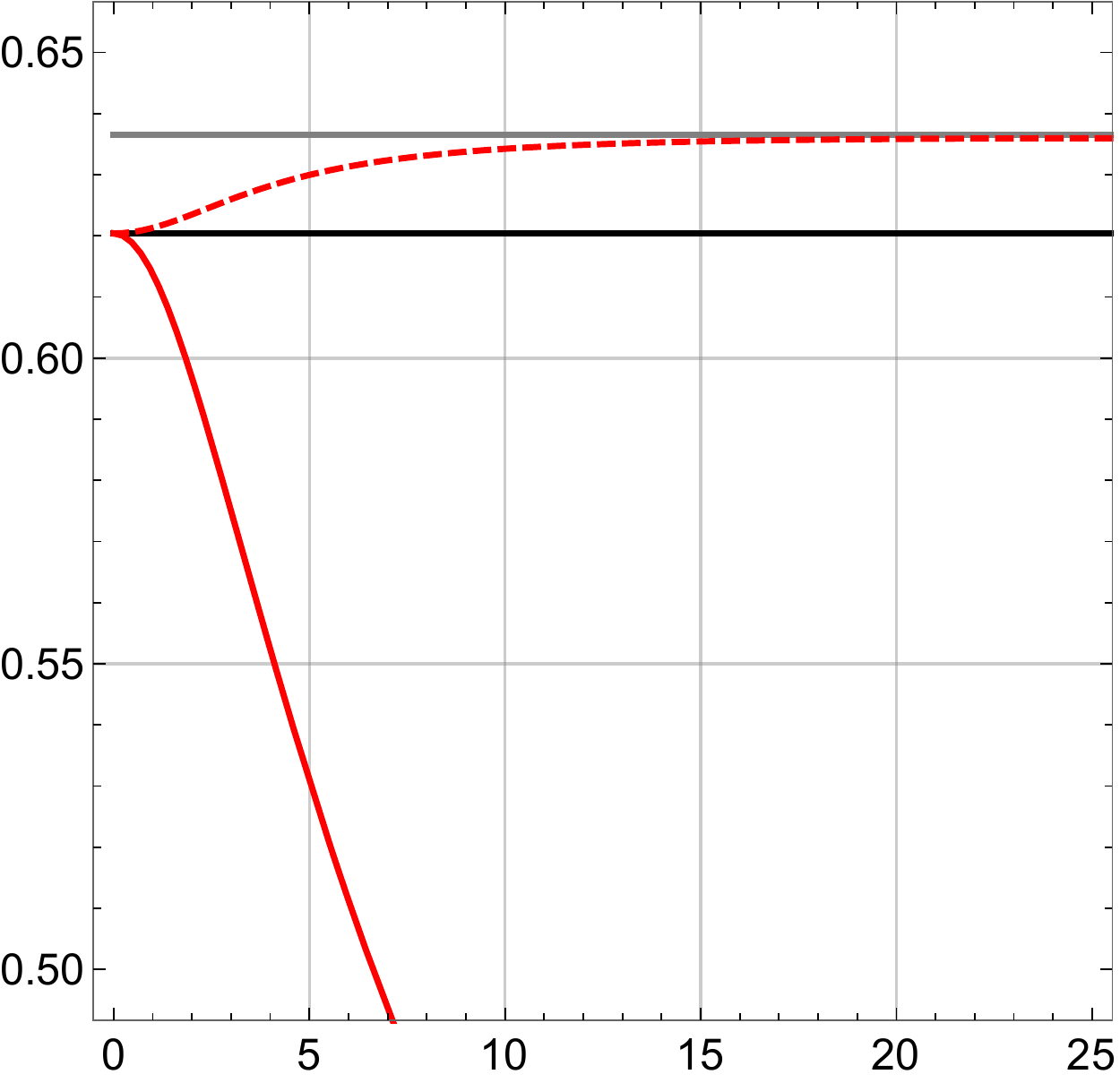}
\qquad
  \put(-425,65){\rotatebox{90}{\large $S^\text{reg}(A \cup B)$}}
         \put(-325,-10){\large $\log \alpha$}
         \put(-205,95){\rotatebox{90}{\large $v_\mt{E}$}}
         \put(-87,-10){\large $a/T$}
         \put(-87,-30){$(b)$}
         \put(-315,-30){$(a)$}
         \put(-145,80){\scriptsize $v_\mt{E}^{\perp}$}
         \put(-60,150){\scriptsize $v_\mt{E}^{||}$}
         \put(-145,135){\scriptsize $v_\mt{E}^{\mt{iso}}$}
         \put(-145,165){\scriptsize $v_\mt{E}^{\mt{Lif}}$}
         
\end{tabular}
\end{center}
\caption{ \small (a) $S^{\text{reg}}(A \cup B)$ as a function of $\log \alpha$. Here we consider a strip oriented orthogonally to the anisotropic direction, $a/T=8.56$ and $\uh=1.097$. The dashed line shows the linear behaviour of $S^{\text{reg}}(A \cup B)$ as a function of $\text{log}(\alpha)$. The angular coefficient of this straight line is given by Eq. (\ref{Slog}). (b) Entanglement velocity as a function of the anisotropy parameter. The continuous red curve represent the result for $v_\mt{E}^{\perp}$, while the dashed red curve represents the result for $v_\mt{E}^{||}$. The black horizontal line is the isotropic result $v_\mt{E}^{\mt{iso}}=\sqrt{2}/3^{3/4}$, whereas the gray horizontal line is the result for a five-dimensional Lifshitz-like geometry, $v_\mt{E}^{\mt{Lif}}=\frac{5^{5/22}\sqrt{11}}{4 \times 2^{10/11}}$.}
\label{Alog-VE}
\end{figure}
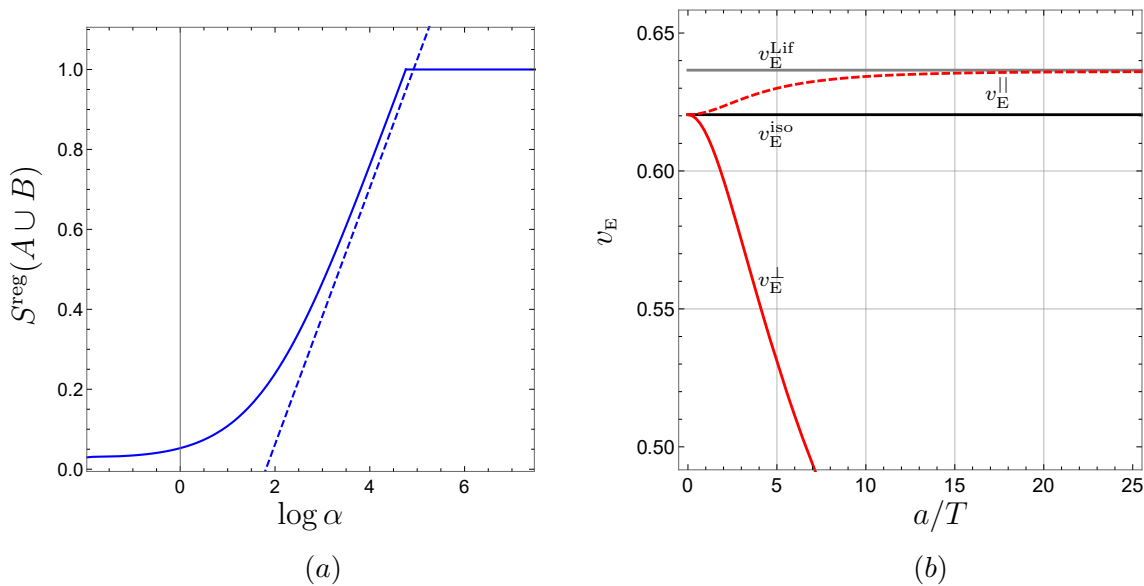

In this section we study the behaviour of $S^{\text{reg}}(A \cup B)$ as a function of the shock wave time $t_0$. In particular, we show that the behavior of $S^\text{reg}(A \cup B)$ as a function of $t_0$ is very similar to the time behaviour of the entanglement entropy of large subregions in holographic models of global quenches \cite{HM,tsunami1,tsunami2}. Let us first consider the case of a semi-infinite strip orthogonal to the anisotropic direction. The function $\alpha_{\perp}(u_0)$ increases as we move $u_0$ deeper behind the horizon and diverges at some point $u_0=u_c^{\perp}$. In the vicinity of $u_c^{\perp}$, we can show that\footnote{This limit was also studied in \cite{Leichenauer-2014}.}
\be
S^\text{reg}(A \cup B) \cong \,\frac{V_2}{G_\mt{N}} \,|G_{xx}(u_0)|\,\sqrt{-G_{tt}(u_0)} \,\frac{\beta}{4 \pi}\, \text{log}\,\alpha\,, \,\,\,\, \text{for}\,\,\,\, u_0 \approx u_c^{\perp} \,.
\label{Slog}
\ee
In figure \ref{Alog-VE} (a), we plot $S^{\text{reg}}(A \cup B)$ versus $\log \alpha$ for the MT model and show that the linear behavior given by Eq. (\ref{Slog}) is correct for $1 \lesssim \alpha \leq \alpha^{*}$, where $\alpha^*$ is the value of $\alpha$ at which $S^\text{reg}(A \cup B)$ becomes constant and $I(A,B)=0$.

As the shift $\alpha$ grows exponentially with time, $\alpha = \text{constant} \times e^{2\pi t_0/\beta}$, the above result implies that $S^\text{reg}(A \cup B)$ grows linearly with $t_0$. Using the formula for the thermal entropy density
\be
s=\frac{\sqrt{G_{xx}^2 (\uh) G_{zz}(\uh)}}{4 G_\mt{N}}\,,
\ee
we can eliminate $G_\mt{N}$ in the above equation and write
\be
\frac{d}{d t_0}S^\text{reg}(A \cup B) = 2\, V_2\, s \, \left( \frac{|G_{xx}(u_0)|\,\sqrt{-G_{tt}(u_0)}}{\sqrt{G_{xx}^2 (\uh) G_{zz}(\uh)}} \right)\,.
\ee
In analogy with \cite{HM,tsunami1,tsunami2}, we can define the {\it entanglement velocity} $v_\mt{E}$ as 
\be
v_\mt{E}^{\perp} = \frac{|G_{xx}(u_c^{\perp})|\,\sqrt{-G_{tt}(u_c^{\perp})}}{\sqrt{G_{xx}^2 (\uh) G_{zz}(\uh)}}\,.
\label{eq-VEperp}
\ee
For an isotropic black-brane solution we have $G_{xx}=G_{zz}=1/u^2$, $G_{tt}=(1-u^4/\uh^4)/u^2$ and $u_c/\uh=3^{1/4}$, and the above formula gives $v_\mt{E}=2^{1/2}/3^{3/4}$. This is consistent with the formula obtained by Hartman and Maldacena \cite{HM}
\be
v_\mt{E} =\frac{\sqrt{d}(d-2)^{1/2-1/d}}{[2(d-1)]^{1-1/d}}
\ee
for $AdS_{d+1}$ black brane solutions.

The corresponding entanglement velocity for a semi-infinite strip oriented along the anisotropic direction is given by
\be
v_\mt{E}^{||} = \frac{\sqrt{G_{xx}(u_c^{||})G_{zz}(u_c^{||})}\,\sqrt{-G_{tt}(u_c^{||})}}{\sqrt{G_{xx}^2 (\uh) G_{zz}(\uh)}}\,,
\label{eq-VEpara}
\ee
where $u_0 = u_c^{||}$ is the point at which $\alpha_{||}(u_0)$ diverges (see Eq. (\ref{Eq-ucPara})). In figure \ref{Alog-VE} (b) we plot the entanglement velocities $v_\mt{E}^{\perp}$ and $v_\mt{E}^{||}$ as a function of the anisotropy parameter for the MT model.

The figure  \ref{Alog-VE} (a) shows that, whenever $\alpha \gtrsim 1$ or, equivalently, after a scrambling time $t_0 \gtrsim t_*$, the regularized entanglement entropy $S^{\text{reg}}(A \cup B)$ grows linearly with $t_0$, and this linear behaviour is characterized by the entanglement velocity $v_\mt{E}$. The linear behaviour of $S^{\text{reg}}(A \cup B)$ persists up to a later time, when this quantity has a sharp transition to a constant thermal value. We say that this is a thermal value because the portion of the U-shaped surface that computes $S^{\text{reg}}(A \cup B)$ is very close to the black brane horizon.

The behaviour of $S^{\text{reg}}(A \cup B)$ as a function of the shock wave time $t_0$ is very similar to the time behaviour of the entanglement entropy of large subregions in holographic models of global quenches like\footnote{Here we use the same terminology used in \cite{Mezei-2016} to denote the holographic quench models of Hartmann and Maldacena \cite{HM} and Liu and Suh \cite{tsunami1,tsunami2}.} the {\it end of the world brane model} of \cite{HM} or the {\it Vaidya model} of \cite{tsunami1,tsunami2,Mezei-2016}. The technical reason for this similarity is that the linear growth of the entanglement entropy with time in all these holographic models is due to a piece of the extremal surface that is very close to the critical surface $u=u_c$, the so-called Hartmann-Maldacena surface. This piece of the extremal surface lies inside the black brane horizon and this region is basically the same in the three cases.

Physically, the similarity can be understood as follows. In the Vaidya quench model, for example, one considers a shock wave at $t=0$ and computes how the entanglement entropy of a large subregion of the boundary evolves in time. In this paper, we consider how the area of an extremal surface at $t=0$ (boundary time) changes as we move the shock wave time $t_0$ further into the past. In both cases the value of the entanglement entropy only depends on how much the shock wave is in the past of the extremal surface. As we move the shock wave further into the past, a piece of the extremal surface inside the black brane horizon becomes closer and closer to the critical surface $u=u_c$, and this region of the geometry is responsible for linear behaviour of the entanglement entropy with time.

The linear behavior of $S^\text{reg}(A \cup B)$ with $\log \alpha$ implies that the mutual information $I(A,B;\alpha)$ decreases linearly with $t_0$, and the entanglement velocity plays an important role in this phenomenon. This linear decrease of the mutual information controlled by the entanglement velocity was also observed in \cite{Hosur-2015} in the context of chaos in quantum channels. The general relation between the mutual information and the entanglement velocity was studied in \cite{Casini-2015}, where the positivity of the mutual information was used to prove that the entanglement velocity is bounded by the velocity of light.

\section{Discussion} \label{sec-4}
We have studied the disruption of the two-sided mutual information in anisotropic shock wave geometries. In particular, we have shown that the entanglement velocity plays an important hole in this phenomenon. From the shock wave profile, we extracted several chaos-related properties of this system, namely, the butterfly velocity, the scrambling time, and the Lyapunov exponent.

We find that the Lyapunov exponent saturates the chaos bound, $\lambda_L =2\pi/\beta$, as expected on general grounds \cite{bound-chaos}, whereas the leading order contribution to the scrambling time scales logarithmically with the black brane entropy, $t_* = \frac{2\pi}{\beta}\log S_{\mt{BH}}$.

Figure~\ref{fig-MIversusL} shows the mutual information in the unperturbed geometry as a function of the strip's width $\ell$ for some values of the anisotropy parameter and for strips orthogonal and parallel to the anisotropic direction. We observe that the mutual information for orthogonal strips are always bigger than the corresponding quantity for parallel strips. Moreover, we also observe that for orthogonal strips the critical width $\ell_c$, at which the mutual information vanishes, decreases with the anisotropy while the corresponding quantity for parallel strips is not affected by the anisotropy\footnote{The effects of anisotropy on the critical width $\ell_c$ are similar to its effects on the screening length of a quarkonium static potential \cite{Chernicoff-2012,Giataganas-2012,Misobuchi-2015}. Note, however, that the result for a parallel (orthogonal) strip should be compared to the result for a quarkonium oriented orthogonally (parallel) to the anisotropic direction.}. This implies that anisotropic systems can have non-zero local correlations for smaller regions than isotropic systems. In other words, the anisotropy increases the local entanglement between the two sides of the geometry, as compared to an isotropic system with the same temperature.

Figure \ref{uc-alpha} shows the behavior of the shock wave parameter $\alpha$ as a function of the point $u_0$ that specifies a constant-$u$ surface. $\alpha$ is an increasing function of $u_0$, starting from zero at $u_0=\uh$ and diverging at some critical point $u_0 = u_c$. This point increases with the anisotropy and it is bigger for strips parallel to the anisotropic direction. This seems to indicate that the anisotropy allows for Hartman-Maldacena surfaces that explore a bigger region in the black hole interior. The behaviour of $S^\text{reg}(A \cup B)$ and $I(A,B)$ is shown in figure \ref{MI-SAB}(a) and figure \ref{MI-SAB}(b), respectively. The area of the extremal surfaces grows faster with $\alpha$  when we increase the anisotropy, and the results for orthogonal strips are always above the results for parallel strips. As a consequence, the mutual information drops to zero faster as we increase the anisotropy. Therefore, on  one hand, the anisotropy increases the two-sided entanglement, while, on the other hand, it disrupts this entanglement faster, as compared to an isotropic system with the same temperature.

Figure \ref{Alog-VE} (a) shows that, for some range of $\alpha$, the extremal surface area grows linearly with the time $t_0$ at which the system was perturbed. This figure also shows that the approximation given by Eq. (\ref{Slog}) is valid not only for very large $\alpha$ (or $u_0$ very close to $u_c$), but actually in the range $1 \lesssim \alpha \leq \alpha^*$, where $\alpha^*$ is the value of $\alpha$ at which $S^\text{reg}(A \cup B)$ becomes constant and $I(A,B)=0$. As $\alpha \sim 1$ defines the scrambling time, we can say that the linear approximation is valid when $t_0 \gtrsim t_*$ up to a later time. The behaviour of $S^\text{reg}(A \cup B)$ as a function of $t_0$ suggest that the gravitational set up used in this paper can be thought of as another example of a quench protocol, where the quench effectively starts when the shock wave time $t_0$ is larger than the scrambling time $t_*$. It might be interesting to investigate if the above set up can provide further insights on the interplay of chaos and spreading of entanglement in strongly coupled systems.

The results  for the entanglement velocities as functions of the anisotropy are shown in Fig \ref{Alog-VE} (b). While $v_\mt{E}^{\perp}$ decreases with the anisotropy, staying below the isotropic value, $v_\mt{E}^{||}$ increases with the anisotropy and seems to approach a constant value for $a/T>>1$, staying always above the isotropic value. Interestingly, an upper bound for the entanglement velocity was proposed in \cite{tsunami1,tsunami2} and derived in \cite{Mezei-2016}. The derivation of the bound relied on imposing a null energy condition in an isotropic background. The bound depends on the dimensionality of the spacetime, and is usually written in terms of the entanglement velocity calculated for an AdS-Schwarzschild black hole $v_\mt{E} \leq v_\mt{E}^\mt{Sch}$. For five-dimensional spacetimes this bound is given by the isotropic result $v_\mt{E}^{\mt{iso}}$, which is violated by $v_\mt{E}^{||}$. This is not in contradiction with \cite{tsunami1,tsunami2,Mezei-2016} because these papers assume isotropy. A bound was also derived for the butterfly velocity \cite{Mezei-2016}, and is given by $v_\mt{B} \leq v_\mt{B}^\mt{Sch}$. For a five-dimensional spacetime, this bound is given by $v_\mt{B}^\mt{iso}$.

Finally, we comment on the results for the butterfly velocity. We first observe that our formulas for the butterfly velocities $v_\mt{B}^{\perp}$ and $v_\mt{B}^{||}$ in generic anisotropic backgrounds (see Eqs. (\ref{VBpara}) and (\ref{VBperp})) agree with previously reported results \cite{Blake1,Wu-2017,Ling-2016,Giataganas-2017}. The specialization of our formulas to the MT model is shown in figure \ref{fig-vB}. The butterfly velocity along the anisotropic direction $v_\mt{B}^{||}$ decreases with the anisotropy, staying below the isotropic value, whereas $v_\mt{B}^{\perp}$ increases with the anisotropy and approaches a constant value for $a/T >>1$, staying always above the bound given by the isotropic result. Again, this violation is not in contradiction with \cite{Mezei-2016} because that paper assumes isotropy.

The behaviour of the butterfly and the entanglement velocity can both be explained by the fact that the MT geometry can be viewed as a renormalization group (RG) flow from an AdS geometry in the ultraviolet (UV) to a Lifshitz-like geometry in the infrared (IR). The parameter that controls this transition is the ratio $a/T$, which is small in the UV and large in the IR. When $a/T$ is small, the geometry is asymptotically AdS and we expect the values of the butterfly and the entanglement velocities to be very close to the corresponding conformal values, which are $v_\mt{B}^{\mt{iso}\,2}=2/3$ 
and $v_\mt{E}^{\mt{iso}}=\sqrt{2}/3^{3/4}$, respectively. For $a/T >>1$, we expect the butterfly and the entanglement velocities to be both given by the effective IR Lifshitz theory, which gives (see appendix \ref{appC})
\be
v_\mt{B}^{||\,2}= \frac{11}{16}r_{\mt{H}}^{2/3}, \,\,\,\,\,\,v_\mt{B}^{\perp\,2}=\frac{11}{16} \equiv v_\mt{B}^{\mt{Lif}\,2}\,,
\ee 
for the butterfly velocities and
\be
v_\mt{E}^{||}= \frac{5^{5/22} \sqrt{11}}{4 \times 2^{10/11}} \equiv v_\mt{E}^{\mt{Lif}}   , \,\,\,\,\,\,v_\mt{E}^{\perp}=\frac{7^{7/22} \sqrt{11}\, r_{\mt{H}}^{1/3}}{3 \times 2^{9/11} \times 3^{7/11}}\,,
\ee
for the entanglement velocities. The velocities $v_\mt{B}^{||}$ and $v_\mt{E}^{\perp}$ are both suppressed at low temperatures (or $a/T >> 1$) because they are proportional to $r_{\mt{H}}^{1/3} \sim T^{1/3}$, while $v_\mt{B}^{\perp}$ and $v_\mt{E}^{||}$ remain constant. \footnote{The scaling of $v_\mt{B}^{||}$ and $v_\mt{E}^{\perp}$ with $r_\mt{H}$ is explained in appendix \ref{appC}.} The results of figure \ref{fig-vB} and figure \ref{Alog-VE} (b) show that the butterfly and the entanglement velocities interpolate between the IR and the UV values, diagnosing the corresponding RG flow. Both velocities respect the  bounds $v_\mt{B} \leq 1$ and $v_\mt{E} \leq 1$, as required for the micro-causality of the UV theory.

As the bounds for $v_\mt{E}$ and $v_\mt{B}$ are derived assuming isotropy and for Einstein gravity, we expect these bounds to be generically violated in anisotropic systems and in higher curvature gravity. Indeed, a violation in the bound for $v_\mt{B}$ for was recently reported in \cite{Giataganas-2017} for an anisotropic and confining system. We believe, however, that these velocities should still be bounded by their corresponding values in the IR effective theory, as it happens in the MT model.

Possible extensions of this work include the study of the chaotic properties of others anisotropic backgrounds and, more generally, of higher curvature gravity theories. Some works in this direction include, for instance, \cite{Blake1,Blake2,Wu-2017,Ling-2016,Roberts-2016,Qaemmaqami-2017,Qaemmaqami-2017-2, Ahn-2017}.


\subsection*{Acknowledgements}
It is a pleasure to thank Diego Trancanelli and Anderson Misobuchi for helpful discussions and insightful comments on the draft. We also thank Juan Pedraza, Dimitrious Giataganas and M\'ark Mezei for useful correspondence. We are indebted to an anonymous referee for helpful suggestions and comments. Finally, we thank Cristiane Jahnke for pointing out some typos in the first version of this paper. This work was supported by Mexico's National Council of Science and Technology (CONACyT) grant CB-2014/238734.

\appendix

\section{Relation between $\alpha$ and $u_0$} \label{appA}

In this appendix we determine the relation between the shock wave parameter $\alpha$ and the point $u_0$ used to compute extremal areas in the shock wave geometry. This point lies behind the horizon in a constante-$u$ surface. By symmetry, the extremal surface homologous to $A \medcup B$ divides the bulk into two halves, as shown in figure \ref{fig-surfaceLocation}. Following \cite{Leichenauer-2014}, we split the left of the surface into three segments $I$, $II$ and $III$. The first segment connects the boundary $(U,V)=(1,-1)$ to the horizon $(U,V)=(U_1,0)$. The second segment goes from the horizon $(U,V)=(U_1,0)$ to the point $(U,V)=(U_2,V_2)$ where the extremal surface intersects with the Hartman-Maldacena surface. In Poincare coordinates this point is specified by $u=u_0$ and some $t$. The third segment connects the point $(U,V)=(U_2,V_2)$ to the horizon at $(U,V)=(0,\alpha/2)$. In what follows we compute the unknown quantities $U_1, U_2$ and $V_2$ in terms of $u_0$ and obtain an expression for $\alpha(u_0)$. For convenience, we remember the definition of Kruskal coordinates\footnote{These are the Kruskal coordinates in the left exterior region of the geometry. Inside the black hole, for example, these coordinates are defined as $U=e^{\frac{2\pi}{\beta}(u_*-t)}$,  $V=e^{\frac{2\pi}{\beta}(u_*+t)}$ and $u_{*}=-\int_u^{\bar{u}} du' \sqrt{\frac{G_{uu}}{G_{tt}}}$.}
\be
U=e^{\frac{2\pi}{\beta}(u_*-t)}\,, \,\,\,\, V=-e^{\frac{2\pi}{\beta}(u_*+t)}\,,\,\,\,\,u_{*}=-\int_0^{u} du' \sqrt{\frac{G_{uu}(u')}{G_{tt}(u')}}\,.
\ee
We consider first the case of a strip oriented orthogonally to the anisotropic direction. From Eq. (\ref{eq-gammaPerp}) the time $t_{\perp}(u)$ along the extremal surface can be written as 
\be
t_{\perp}(u)=\int du\,\sqrt{\frac{G_{uu}}{G_{tt}}} \frac{1}{\sqrt{1+G_{xx}^2 G_{tt} \gamma_{\perp}^{-2}}}\,.
\ee
Using the above equations we can express variation in the coordinates $U$ and $V$ as
\bea
&\Delta \log U^2 &= \frac{4\pi}{\beta}\left(\Delta u_*-\Delta t \right)=\frac{4\pi}{\beta} \int du\, \sqrt{\frac{G_{uu}}{G_{tt}}}\left(\frac{1}{\sqrt{1+G_{xx}^2 G_{tt} \gamma_{\perp}^{-2}}}-1 \right) \,,\cr
&\Delta \log V^2 &= \frac{4\pi}{\beta}\left(\Delta u_*+\Delta t \right)=\frac{4\pi}{\beta} \int du\, \sqrt{\frac{G_{uu}}{G_{tt}}}\left(\frac{1}{\sqrt{1+G_{xx}^2 G_{tt} \gamma_{\perp}^{-2}}}+1\right) \,.
\eea
The coordinate $U_1$ can be calculated considering the variation of $U$ from the boundary to the horizon
\be
U_1^2 = \text{exp}\left[\frac{4\pi}{\beta} \int_{0}^{\uh} du\, \sqrt{\frac{G_{uu}}{G_{tt}}}\left(\frac{1}{\sqrt{1+G_{xx}^2 G_{tt} \gamma_{\perp}^{-2}}}-1 \right) \right]\,.
\ee
To compute $U_2$ we consider the variation of $U$ from $u=\uh$ to $u=u_0$
\be
\frac{U_2^2}{U_1^1} = \text{exp}\left[ \frac{4\pi}{\beta} \int_{\uh}^{u_0} du\,\sqrt{\frac{G_{uu}}{G_{tt}}}\left(\frac{1}{\sqrt{1+G_{xx}^2 G_{tt} \gamma_{\perp}^{-2}}}-1 \right) \right]\,.
\ee
The coordinate $V_2$ can be written as
\be
V_2=\frac{1}{U_2} \, \text{exp}\left[ \frac{4\pi}{\beta} \int_{\bar{u}}^{u_0} du\,\sqrt{\frac{G_{uu}}{G_{tt}}}\right]\,,
\ee
where $\bar{u}$ is a point behind the horizon at which $u_*=0$. The shift $\alpha$ can then be computed by considering the variation in the $V$-coordinate along the segment $III$
\be
\frac{\alpha^2}{4 V_2^2}=\text{exp}\left[\frac{4\pi}{\beta} \int_{u_0}^{\uh} du\,\sqrt{\frac{G_{uu}}{G_{tt}}}\left(\frac{1}{\sqrt{1+G_{xx}^2 G_{tt} \gamma_{\perp}^{-2}}} -1\right) \right]=\frac{U_1^2}{U_2^2}\,.
\ee
After some simplifications, the parameter $\alpha$ can be written as
\be
\alpha_{\perp}(u_0)=2\, e^{K_1^{\perp}(u_0)+K_2^{\perp}(u_0)+K_3^{\perp}(u_0)}\,,
\ee
where
\bea
&K_1^{\perp}&= -\frac{4 \pi}{\beta}  \int_{\bar{u}}^{u_0} du\, \sqrt{\frac{G_{uu}}{G_{tt}}}\,,\cr
&K_2^{\perp}&=\frac{2 \pi}{\beta}  \int_{0}^{u_\mt{H}} du\, \sqrt{\frac{G_{uu}}{G_{tt}}} \left(1-\frac{1}{\sqrt{1+G_{xx}^2 G_{tt} \gamma_{\perp}^{-2}}} \right)\,,\cr
&K_3^{\perp}&=- \frac{4 \pi}{\beta} \int_{u_\mt{H}}^{u_0} du\, \sqrt{\frac{G_{uu}}{G_{tt}}} \left(1-\frac{1}{\sqrt{1+G_{xx}^2 G_{tt} \gamma_{\perp}^{-2}}} \right)\,,
\eea
where we use the subscript $\perp$ to indicate that these quantities were calculated for a strip orthogonal to the anisotropic direction. The equivalent expressions for a strip oriented parallel to the anisotropic direction are
\be
\alpha_{||}(u_0)=2\, e^{K_1^{||}(u_0)+K_2^{||}(u_0)+K_3^{||}(u_0)}\,,
\ee
where
\bea
&K_1^{||}&=-\frac{4 \pi}{\beta} \int_{\bar{u}}^{u_0} du\, \sqrt{\frac{G_{uu}}{G_{tt}}}\,,\cr
&K_2^{||}&= \frac{2 \pi}{\beta} \int_{0}^{u_\mt{H}} du\, \sqrt{\frac{G_{uu}}{G_{tt}}} \left(1-\frac{1}{\sqrt{1+G_{xx} G_{zz} G_{tt} \gamma_{||}^{-2}}} \right)\,,\cr
&K_3^{||}&=-\frac{4 \pi}{\beta} \int_{u_\mt{H}}^{u_0} du\, \sqrt{\frac{G_{uu}}{G_{tt}}} \left(1-\frac{1}{\sqrt{1+G_{xx} G_{zz} G_{tt} \gamma_{||}^{-2}}} \right)\,.
\eea

\section{Divergence of $K_3(u_0)$} \label{appB}

In this section we determine the critical point $u_0 =u_c$ at which $K_3(u_0)$ diverges. $K_3$ is given by
\be
K_{3}^{||}=- 4 \pi T \int_{u_\mt{H}}^{u_0} du \sqrt{\frac{G_{uu}}{G_{tt}}} \left(1-\frac{1}{\sqrt{1+G_{xx}G_{zz} G_{tt} \gamma_{\perp}^{-2}}} \right)
\label{eq-K3para}
\ee
for a strip orthogonal to the anisotropic direction and by
\be
K_{3}^{\perp}=- 4 \pi T \int_{u_\mt{H}}^{u_0} du \sqrt{\frac{G_{uu}}{G_{tt}}} \left(1-\frac{1}{\sqrt{1+G_{xx}^2 G_{tt} \gamma_{\perp}^{-2}}} \right)
\ee
for a strip oriented along the anisotropic direction. Let us consider first the critical point of $K_3^{||}$, the result for $K_3^{\perp}$ can be obtained from the result for $K_3^{||}$ by replacing $G_{zz}$ by $G_{xx}$.

The critical point $u_c$ can be obtained by considering the integrand of Eq. (\ref{eq-K3para}) in the limit where $u \rightarrow u_0$. Note that
\bea
G_{xx}G_{zz} G_{tt} \gamma_{\perp}^{-2} &=&-\frac{G_{xx}G_{zz} G_{tt}}{G_{xx}(u_0)G_{zz}(u_0) G_{tt}(u_0)} \nonumber\\ 
& = & \frac{G_{xx}(u_0)G_{zz}(u_0) G_{tt}(u_0)+\big( G_{xx}G_{zz} G_{tt}\big)'\big|_{u=u_0} (u-u_0)}{G_{xx}(u_0)G_{zz}(u_0) G_{tt}(u_0)}+\mathcal{O}(u-u_0)^2\nonumber \\ 
&=& 1+\frac{\big( G_{xx}G_{zz} G_{tt}\big)'}{ G_{xx}G_{zz} G_{tt}}\Big|_{u=u_0}(u-u_0)+\mathcal{O}(u-u_0)^2\,.
\eea
Using the above result in Eq. (\ref{eq-K3para}) one finds
\be
K_{3}^{||}\approx -4 \pi T \int_{u_\mt{H}}^{u_0} du \sqrt{\frac{G_{uu}(u_0)}{G_{tt}(u_0)}} \left(1-\frac{1}{\sqrt{-\frac{\big( G_{xx}G_{zz} G_{tt}\big)'}{ G_{xx}G_{zz} G_{tt}}\Big|_{u=u_0}(u-u_0)}} \right)\,.
\ee
The above expression diverges when $u_0 \rightarrow u_c^{||}$ such that
\be
\frac{\big(G_{xx}G_{zz} G_{tt}\big)'}{ G_{xx}G_{zz} G_{tt}}\Big|_{u=u_c^{||}}=0\,.
\ee
The corresponding expression for $u_c^{\perp}$ (the point where $K_3^{\perp}$ diverges) is
\be
\frac{\big( G_{xx}^2 G_{tt}\big)'}{ G_{xx}^2 G_{tt}}\Big|_{u=u_c^{\perp}}=0\,.
\ee
As a first check for these expressions, let us consider the case at which $G_{xx}=G_{zz}=r^2$, and $G_{tt}=f(r)$. The equation for $r_c$ is
\be
\frac{\big(r^4 f(r)\big)'}{r^4 f(r)} \Big|_{u=u_c} = \frac{4}{r_c}+\frac{f'(r_c)}{f(r_c)}=0\,.
\ee
Multiplying the above result by $f(r_c) r_c$ we obtain the Eq. (40) of \cite{Leichenauer-2014} for $d=4$, as expected.

\section{Anisotropic black branes: the MT model} \label{appC}

In this section we briefly review the anisotropic black brane solution of Mateos and Trancanelli \cite{MT1,MT2}. This is a solution of type IIB supergravity whose metric in the Einstein frame reads
\be
ds^2 = \frac{\ell_\text{AdS}^2 e^{-\frac{\phi(u)}{2}}}{u^2} \left(-B(u)F(u) dt^2+dx^2+dy^2+H(u)dz^2+\frac{du^2}{F(u)} \right)+L^2 d\Omega_5^2\,,
\label{MTsol}
\ee
where $H(u)=e^{-\phi(u)}$ and $\Omega_5$ is the volume form of a round 5-sphere. The AdS radial coordinate is $u$ while the boundary theory coordinates are $(t,x,y,z)$. The above metric has a horizon at $u=\uh$ and the boundary is located at $u=0$. We set the AdS radius $\ell_\text{AdS}$ to unity in the following. The effects of the anisotropy on the geometry are controlled by the ratio $a/T$, where $T$ is the black brane Hawking temperature and $a$ is the parameter of anisotropy. 

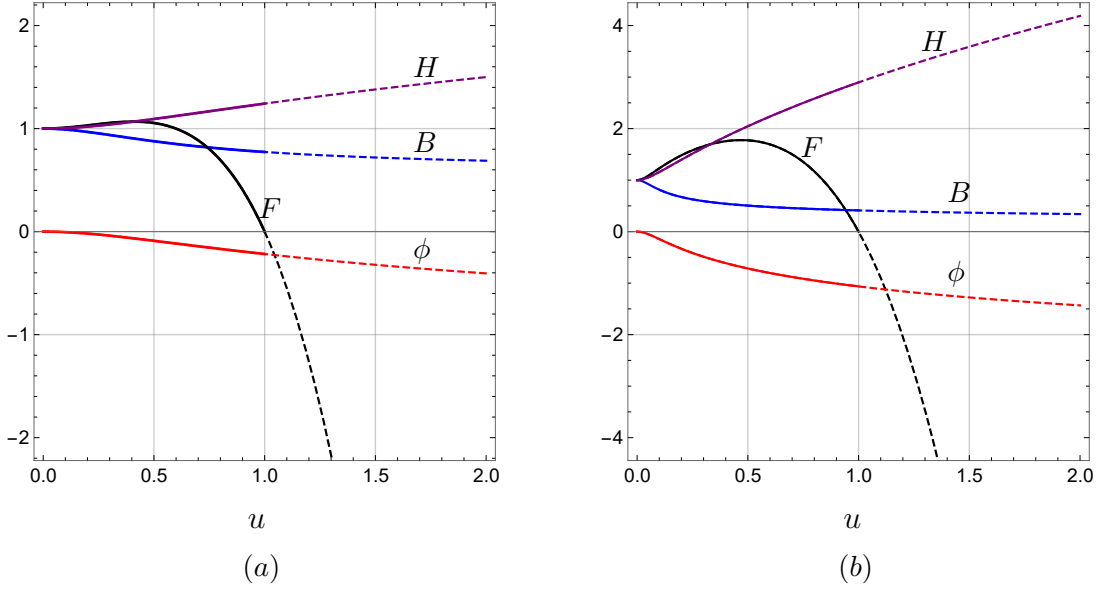
\begin{figure}[H]
\begin{center}
\begin{tabular}{cc}
\setlength{\unitlength}{1cm}
\hspace{-0.9cm}
\includegraphics[width=6.5cm]{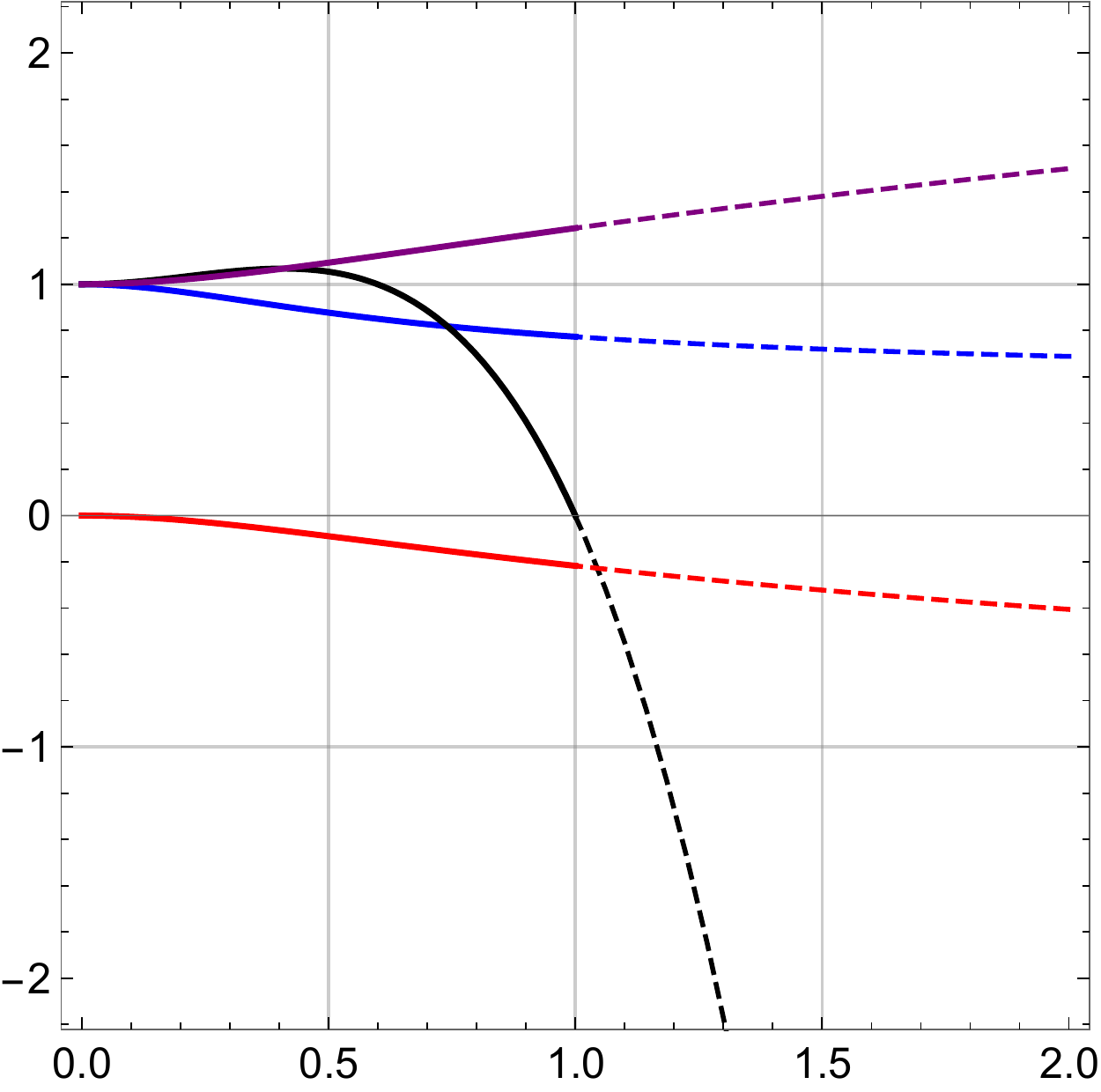} 
\qquad\qquad & 
\includegraphics[width=6.5cm]{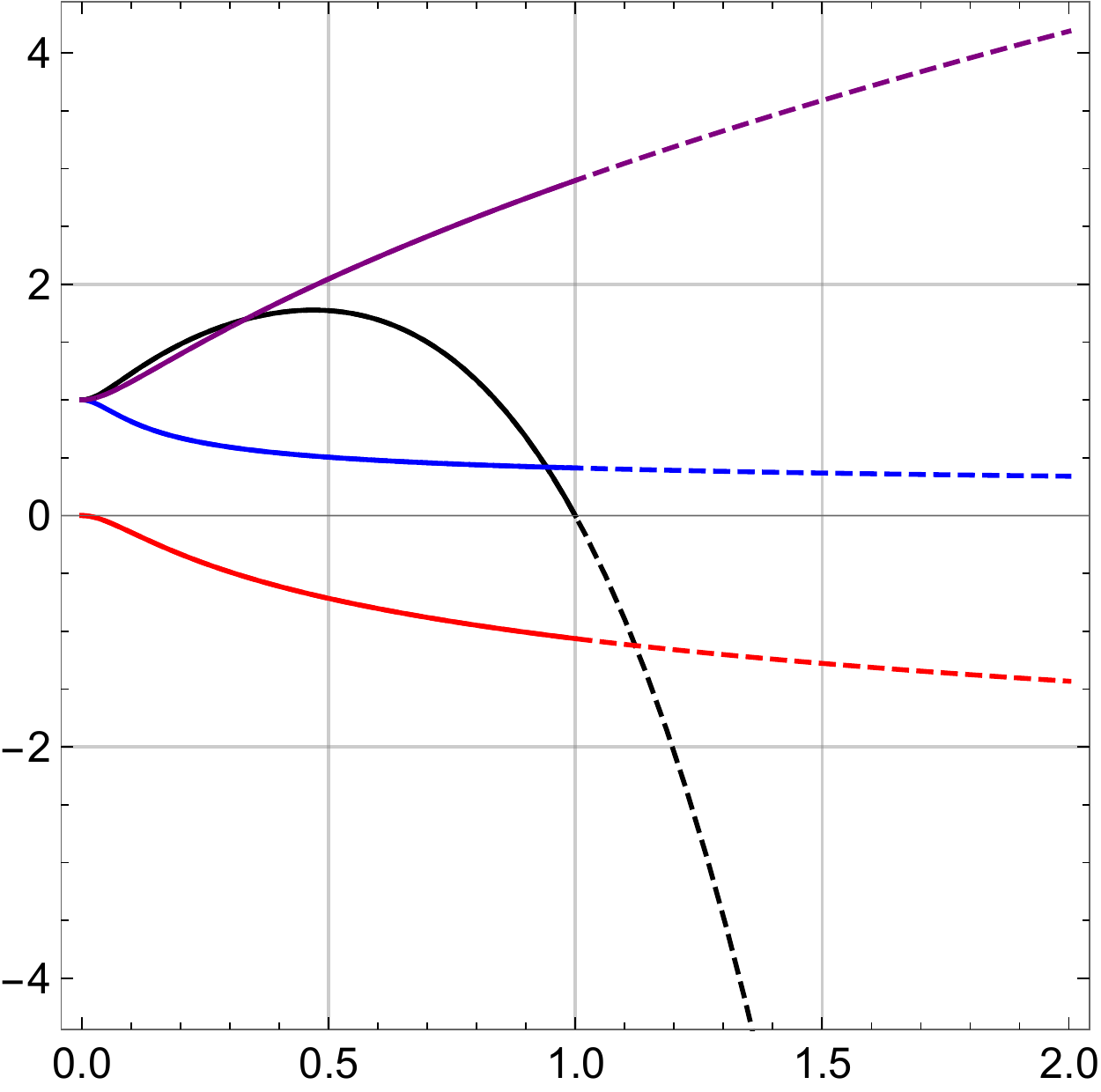}
\qquad
         \put(-317,-15){\large $u$}
         \put(-94,-15){\large $u$}
         \put(-96,-33){$(b)$}
         \put(-319,-33){$(a)$}
         \put(-255,155){$H$}
         \put(-255,127){$B$}
         \put(-255,87){$\phi$}
         \put(-313,102){$F$}
         \put(-65,165){$H$}
         \put(-55,108){$B$}
         \put(-55,78){$\phi$}
         \put(-110,125){$F$}
\end{tabular}
\end{center}
\caption{ \small (a) Metric functions for $a/T \simeq 4.4$. (b) Metric functions for $a/T \simeq 25$. The continuous curves represent the result outside the horizon ($u<\uh$), whereas the dashed curves represent the results inside the horizon ($u>\uh$). Here we have fixed $\uh=1$.}
\label{fig-metric}
\end{figure}

The above solution can be thought of as describing a renormalization group (RG) flow from an AdS geometry in the ultraviolet (UV) to a Lifshitz-like geometry in the infrared (IR). The transition is controlled by the ratio $a/T$, which is small in the UV and large in the IR. The metric functions $B$, $F$, $H$ and $\phi$ are known analytically in the limits of $a/T << 1$ (UV) and $a/T >> 1$ (IR), and numerically in intermediate regimes \cite{MT2}. Figure \ref{fig-metric} shows the behaviour of $B$, $F$, $H$ and $\phi$ as a function of the AdS radial coordinate $u$. This coordinate extends from the boundary $u=0$ up to some point inside the black brane horizon $u=2\uh$. From figure \ref{fig-metric} we can see that, at least for $\uh \leq u \leq 2 \uh$, the metric functions are well-behaved even inside the black brane horizon. 

Finally, we comment that the above solution satisfies the conditions given in Eq. (\ref{cond-sw}) for the existence of consistent shock wave solutions.

\subsection{High-temperatures or small anisotropies ($a/T << 1$)}

In the limit $a/T << 1$ there is an analytic solution for metric functions
\bea
&F(u)&=1-\frac{u^4}{\uh^4}+a^2 F_2(u)\,,\cr
&B(u)&=1+a^2 B_2(u)\,,\cr
&\phi(u)&= a^2 \phi_2(u)\,,
\eea
with
\bea
&F_2(u)&=\frac{1}{24 \uh^2}\left[ \left(7 u^4+3 \uh^4 \right) \log \left(\frac{u^2}{\uh^2}+1\right)-10 u^4 \log 2+8 u^2 \left(\uh^2-u^2\right) \right]\,,\cr
&B_2(u)&= \frac{\uh^2}{24} \left(\frac{10 u^2}{u^2+\uh^2}+\log \left(\frac{u^2}{\uh^2}+1\right)\right)\,,\cr
&\phi_2(u)& = -\frac{\uh^2}{4} \log \left(\frac{u^2}{\uh^2}+1\right)\,.
\eea
The Hawking temperature and the entropy density of the above solution are given by
\bea
T=\frac{1}{\pi \uh}+\frac{\uh (5 \log 2-2)}{48 \pi} a^2\,,\qquad
s= \frac{\pi^2 N_c^2 T^3}{2}+\frac{N_c T}{16}a^2 \,.
\eea
The butterfly velocity along the anisotropic direction and perpendicular to it can be written as
\be
v_\mt{B}^{||\,2}=\frac{2}{3}-\frac{(\uh a)^2}{72}(\log 256+2), \,\,\,\,\,\,v_\mt{B}^{\perp\,2}=\frac{2}{3}+\frac{(\uh a)^2}{72}(\log 16-2)\,.
\ee 
The effect of anisotropy is to increase $v_\mt{B}^{\perp}$ and decrease $v_\mt{B}^{||}$ with respect to the isotropic value $v_\mt{B}^{\mt{iso}}=\sqrt{2/3}$.

The entanglement velocity for parallel and orthogonal strips can be written as
\bea
\nonumber
&v_\mt{E}^{||\,2}&=\frac{2}{3\sqrt{3}}-\frac{(\uh a)^2}{108}\left(3\sqrt{3}+4\sqrt{3}\log[1+\sqrt{3}]-3\right)\,,\\
&v_\mt{E}^{\perp\,2}&=\frac{2}{3\sqrt{3}}+\frac{(\uh a)^2}{108}\left(3-3\sqrt{3}+2\sqrt{3}\log[1+\sqrt{3}]\right)\,.
\eea 
In this case the effect of the anisotropy is to increase $v_\mt{E}^{||}$ and decrease $v_\mt{E}^{\perp}$ with respect to the isotropic value $v_\mt{E}^{\mt{iso}}=\sqrt{2}/3^{3/4}$.

\subsection{Low-temperatures or high anisotropies ($a/T >> 1$)}
In the limit $a/T >> 1$ the solution (\ref{MTsol}) flows to a Lifshitz-like geometry \cite{MT2}, whose metric in Einstein frame reads \cite{Azeyanagi-2009}
\be
ds^2=L^2 \left[r^2(-F(r)dt^2+dx^2+dy^2)+r^{4/3}dz^2+\frac{dr^2}{r^2 F(r)} \right]+L^2 d \Omega_5^2\,.
\label{metric-Lif}
\ee
with
\be
F(r)=1-\left( \frac{r_{\mt{H}}}{r} \right)^{11/3}\,,
\ee
where $r_{\mt{H}}$ is the position of the horizon and the coordinate $r$ is related to the $u$-coordinate as $r^{7/3} \sim u^{-2}$ \cite{MT2}. The Hawking temperature and the entropy density are given by
\be
T = \frac{11 r_{\mt{H}}}{12 \pi}, \,\,\,\,\,\, s = \text{constant} \times N_c^2\, a^{1/3}\,T^{8/3}\,.
\ee
The butterfly velocity along the anisotropic direction and perpendicular to it can be written as
\be
v_\mt{B}^{||\,2}= \frac{11}{16}r_{\mt{H}}^{2/3}, \,\,\,\,\,\,v_\mt{B}^{\perp\,2}=\frac{11}{16} \equiv v_\mt{B}^{\mt{Lif}\,2} \,,
\ee
where the scaling $v_\mt{B}^{||} \sim r_{\mt{H}}^{1/3}$ is due to the fact that the coordinate $z$ in the metric (\ref{metric-Lif}) has dimension 2/3, while the time coordinate $t$ has dimension 1, and so the dimension of the butterfly velocity in this direction is $[dz/dt]=1/3$. 
The entanglement velocity for parallel and orthogonal strips reads
\be
v_\mt{E}^{||}= \frac{5^{5/22} \sqrt{11}}{4 \times 2^{10/11}} \equiv v_\mt{E}^{\mt{Lif}\,2}   , \,\,\,\,\,\,v_\mt{E}^{\perp}=\frac{7^{7/22} \sqrt{11}\, r_{\mt{H}}^{1/3}}{3 \times 2^{9/11} \times 3^{7/11}}\,,
\ee 
where the scaling $v_\mt{E}^{\perp} \sim r_{\mt{H}}^{1/3}$ is due to the presence of a factor of $G_{zz}^{1/2}$ in the denominator of the formula for $v_\mt{E}^{\perp}$ and no such factor in the numerator. See Eqs. (\ref{eq-VEperp}) and (\ref{eq-VEpara}).
The velocities $v_\mt{B}^{||}$ and $v_\mt{E}^{\perp}$ are both suppressed at low temperatures because they are proportional to $r_{\mt{H}}^{1/3} \sim T^{1/3}$, while $v_\mt{B}^{\perp}$ and $v_\mt{E}^{||}$ remain constant. In the main text we show that $v_\mt{B}$ and $v_\mt{E}$ interpolate smoothly between the UV and the IR values as we increase the ratio $a/T$.



\end{document}